\documentclass[acmlarge]{acmart}

\usepackage{enumitem} %
\usepackage{multirow} %
\usepackage{subfig} %

\AtBeginDocument{%
  \providecommand\BibTeX{{%
    \normalfont B\kern-0.5em{\scshape i\kern-0.25em b}\kern-0.8em\TeX}}}

\setcopyright{acmlicensed}
\acmJournal{IMWUT}
\acmYear{2023} \acmVolume{7} \acmNumber{2} \acmArticle{69} \acmMonth{6} \acmPrice{15.00}\acmDOI{10.1145/3596268}

\begin{document}

\title{BMAR: Barometric and Motion-based Alignment and Refinement for Offline Signal Synchronization across Devices}

\author{Manuel Meier}
\email{manuel.meier@inf.ethz.ch}
\orcid{0000-0001-6593-7695}
\affiliation{%
  \institution{Department of Computer Science, ETH Z{\"u}rich}
  \country{Switzerland}
}

\author{Christian Holz}
\email{christian.holz@inf.ethz.ch}
\orcid{0000-0001-9655-9519}
\affiliation{%
  \institution{Department of Computer Science, ETH Z{\"u}rich}
  \country{Switzerland}.
}

\renewcommand{\shortauthors}{Meier and Holz}

\begin{abstract}

A requirement of cross-modal signal processing is accurate signal alignment.
Though simple on a single device, accurate signal synchronization becomes challenging as soon as multiple devices are involved, such as during activity monitoring, health tracking, or motion capture---particularly outside controlled scenarios where data collection must be standalone, low-power, and support long runtimes.
In this paper, we present \emph{BMAR}, a novel synchronization method that operates purely based on recorded signals and is thus suitable for offline processing.
\emph{BMAR} needs no wireless communication between devices during runtime and does not require any specific user input, action, or behavior.
\emph{BMAR} operates on the data from devices worn by the same person that record  barometric pressure and acceleration---inexpensive, low-power, and thus commonly included sensors in today's wearable devices.
In its first stage, \emph{BMAR} verifies that two recordings were acquired simultaneously and pre-aligns all data traces.
In a second stage, \emph{BMAR} refines the alignment using acceleration measurements while accounting for clock skew between devices.
In our evaluation, three to five body-worn devices recorded signals from the wearer for up to ten hours during a series of activities.
\emph{BMAR} synchronized all signal recordings with a median error of 33.4\,ms and reliably rejected non-overlapping signal traces.
The worst-case activity was sleeping, where \emph{BMAR's} second stage could not exploit motion for refinement and, thus, aligned traces with a median error of 3.06\,s.

\end{abstract}

\begin{CCSXML}
<ccs2012>
   <concept>
       <concept_id>10010583.10010588.10010596</concept_id>
       <concept_desc>Hardware~Sensor devices and platforms</concept_desc>
       <concept_significance>500</concept_significance>
       </concept>
   <concept>
       <concept_id>10010583.10010588.10010595</concept_id>
       <concept_desc>Hardware~Sensor applications and deployments</concept_desc>
       <concept_significance>500</concept_significance>
       </concept>
   <concept>
       <concept_id>10010583.10010588.10003247.10003248</concept_id>
       <concept_desc>Hardware~Digital signal processing</concept_desc>
       <concept_significance>500</concept_significance>
       </concept>
   <concept>
       <concept_id>10010520.10010553.10010562</concept_id>
       <concept_desc>Computer systems organization~Embedded systems</concept_desc>
       <concept_significance>300</concept_significance>
       </concept>
 </ccs2012>
\end{CCSXML}

\ccsdesc[500]{Hardware~Sensor devices and platforms}
\ccsdesc[500]{Hardware~Sensor applications and deployments}
\ccsdesc[500]{Hardware~Digital signal processing}
\ccsdesc[300]{Computer systems organization~Embedded systems}

\keywords{Synchronization, signal processing, wearable devices, embedded systems, monitoring, data analysis.}

\begin{teaserfigure}
    \centering
    \vspace{-2mm}
    \includegraphics[width=.9\textwidth]{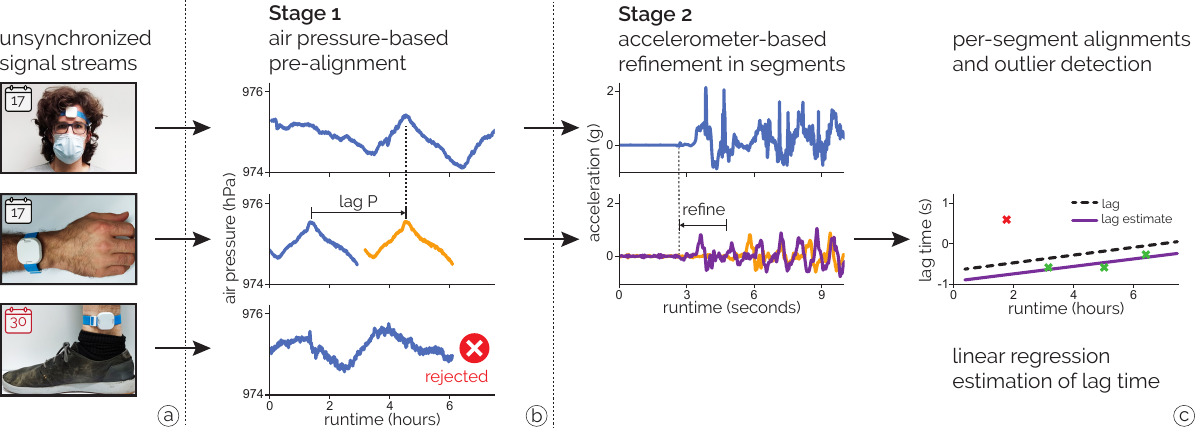}
    \vspace{-2mm}
    \caption{\emph{BMAR} is a novel method to synchronize and align signals across devices without the need for specific user input, action, or explicit synchronization through wired or wireless communication (e.g., WiFi or BLE).
    BMAR is capable of synchronizing (a)~independently recorded signals \emph{after the fact} by (b)~first pre-aligning recordings using air pressure as an inexpensive sensing modality that simultaneously allows us to reject non-simultaneous recordings.
    (c)~In a second step, BMAR produces a refined signal alignment across sensor devices by cross-correlating accelerometer observations.}
    \label{fig:overview}
\end{teaserfigure}

\maketitle

\section{Introduction}

Technological advances have led to a fast increase in the miniaturization of technology, giving rise to many devices that can be worn throughout the day~\cite{bisht_study_2021}.
Among the main uses of wearable devices are activity tracking~\cite{kamisalic2018wearablesactivityhealth, strain2020wearablesactivityhealth, Gjoreski2016wearable, Case2015wearable, Liu2016wearable, dahmen2017wearables, Alinia2017wearables} and health monitoring~\cite{kamisalic2018wearablesactivityhealth, strain2020wearablesactivityhealth, Zheng2014unobtrusive, an2017wearable, Bloss2015wearable, malhi2012zigbee}.
Both these types of devices incorporate multiple sensors that generate large amounts of time-dependent data that they must process in real-time, store, or transmit to another device.
Oftentimes, the data is processed online to reduce storage or transmission requirements.

When raw recordings obtained on wearable devices are important, however, continuous data streaming places a large toll on battery power and storing all data for offline processing is preferable.
This is especially true of data collection experiments that enable later research efforts~\cite{bruno2021battery}.
Beyond the sensor data collected by one device, data collections commonly include multiple wearable devices, either worn by one individual or across individuals, such that the collected records can later be combined for joint analysis~\cite{park_wearables_2017}.
To analyze data series stored across multiple devices, accurate temporal data synchronization is an essential requirement.
Wearable sensing devices usually associate measurements with timestamps, for example, derived from an on-board real-time clock (RTC). %
Since timekeeping typically relies on a crystal oscillator with an accuracy of 10--100\,ppm~\cite{tirado-andres_performance_2019}, the clocks between two devices may deviate by several seconds after hours of runtime, even if they matched at one point.
Repeated synchronization is therefore essential for accurate data processing at a later point.

A common approach to cross-device time synchronization is using a central reference for timekeeping.
The Global Positioning System (GPS) is a popular choice because it synchronizes time at an accuracy of typically 10\,ns~\cite{dana1990role} which exceeds that of an RTC and is only limited by hardware capabilities~\cite{li_methodology_2015}.
However, GPS signals are not reliable in locations with no clear view of the sky and the receiver's power consumption of several milliwatts renders them little suitable for repeated synchronization in low-power applications.
Therefore, many wearables rely on a wireless sensor network (WSN) that implements an online synchronization algorithm.
While there are no established standards in WSN synchronization, multiple commonly used wireless synchronization protocols exist, including the timing-sync protocol for sensor networks (TPSNs)~\cite{ganeriwal_timing-sync_2003}, the flood time synchronization protocol (FTSP)~\cite{maroti_flooding_2004}, and reference broadcast synchronization (RBS)~\cite{elson_fine-grained_2002}. 
However, all these approaches are designed and optimized for accuracy and are not suitable for very low-power applications~\cite{makara_power_2019}.

A recent trend in low-power systems has been in time synchronization that optimizes for power consumption and thus limits communication and the amount of online processing~\cite{bruno2021battery}.
For example, \citeauthor{makara_power_2019}'s synchronization protocol tolerates average time discrepancies of 10\,ms, enabling two individual wearable devices to regularly exchange timestamps through Bluetooth Low Energy (BLE) at an average power consumption of just $133\,\mu W$~\cite{makara_power_2019}. 
In contrast, many other efforts have opted to perform data synchronization \emph{entirely offline} and based on measurement data alone, which is useful for longitudinal in-the-wild collections.
\citeauthor{spilz_novel_2021} and \citeauthor{gilbert_simple_2022}'s methods build on magnetometer recordings to detect a predefined event for clock synchronization~\cite{spilz_novel_2021, gilbert_simple_2022}, leveraging electromagnets built into a tray on which the devices are placed.
In many applications, such as wearable health sensing or collections of activity data, synchronization is particularly needed when devices are worn by the same person to obtain a coherent dataset for each wearer.
For this use case, researchers have also synchronized data series passively, detecting pre-defined events in the signals from inertial measurement units (IMUs) for synchronization, such as clapping~\cite{shabani_automatic_2022} or footsteps~\cite{bannach_automatic_2009}.

Following from previous efforts in this domain, several key challenges exist for passively synchronizing data series offline:
Synchronization must be able to detect whether or not two data series were recorded at the same time as a prerequisite.
Synchronization cannot rely on pre-defined activities or events that are required to be performed by a person one or more times.
Synchronization cannot assume unique events to be present inside all data series that passively afford accurate alignment.
Synchronization must ensure that data series are aligned with a global optimum, particularly in the presence of events that manifest in periodic patterns, such as walking, running, or cycling, and that can lead to locally optimal but overall inaccurate alignment.

To address the challenges above, we propose \emph{BMAR}, a novel offline synchronization method for devices worn by one person.
BMAR requires no explicit input from users and accurately aligns cross-sensor signal traces without requiring users to perform particular motions or activities.

\subsection{Offline Signal Synchronization across Body-worn Devices}

As shown in \autoref{fig:overview}, BMAR synchronizes signal traces individually recorded by wearable devices by aligning them in two stages: a pre-alignment based on air pressure observations and a refinement step that processes accelerometer observations.

In Stage~1, BMAR derives a pre-alignment of signal traces based on air-pressure observations. 
We simultaneously leverage these measurements for the rejection of non-concurrent recordings. 
Both are possible because air pressure recordings are unique over time but near-identical across all devices worn on the same body.
Barometers are inexpensive, small, very low-power, and not affected by motion artifacts, ideally suited for this scenario on wearable devices.

In Stage~2, BMAR refines alignment based on accelerometer measurements.
Building on our pre-alignment to ensure globally optimal alignment, BMAR produces a refined alignment by cross-correlating accelerometer recordings.
We compute these for patches of data, which makes our method more robust to short-term misalignments and accounts for clock skew between devices.
Accelerometers are ubiquitous in wearable devices thanks to the versatility of their recordings, their low cost, and low power consumption. 
The distinct high-frequency events in motion across all on-body locations that these sensors observe are ideal for optimizing BMAR's pre-alignment to reduce the alignment error.

To evaluate the effectiveness of our method, we conducted 10 sessions of data collection on 7 separate days.
In each session, between 3 and 5 wearable devices recorded data from the same participant for up to 10\,hours, resulting in 328 hours of total device runtime.
The participant performed one of 3 activities: office work, cycling, or skiing to include a variety of lower and higher-intensity motions.
In addition, we included sleep as a 4th activity to verify how well our method performs on recordings with little to no motion.

Our results showed that BMAR synchronizes signals with a median error of less than 40\,ms during sessions with activity.
This is well below the period of human activity and physiological signals~\cite{luz_evaluating_2014, choi_photoplethysmography_2017}, setting our method up as a suitable complement to the increasing data monitoring efforts on multi-device platforms.

\subsection{Contributions}
\label{sub:contributions}

We make the following contributions in this paper:

\begin{enumerate}[leftmargin=*]
    \item BMAR, a novel method for signal synchronization that processes recordings offline through a pre-alignment and refinement stage, without relying on assumptions about signal content (e.g., specific actions or user input during recording).
    
    \item Particular support for (wearable) low-power platforms, requiring no wireless connectivity or communication during operation, and leveraging only commonplace air pressure and accelerometer measurements.
    
    \item An evaluation of our method across 3--5 custom-built wearable devices (with a high-resolution and a low-resolution sensor) across 4 activities against ground-truth measurements.
    During activities with human motion, our method aligned data traces with a median error of 33.4\,ms, whereas this error was 3\,s during activities without motion (sleep).
    Our method also reliably detected and rejected recordings that were not acquired concurrently.
\end{enumerate}

\section{Related Work}

BMAR is related to wireless synchronization methods, data-based synchronization, and wearable applications.

\subsection{Wireless Synchronization}
\label{subsec:rw_wireless_sync}
Over recent years, Wireless sensor networks (WSNs) have attracted more and more interest.
Time synchronization performed online or offline between the nodes of a WSN is crucial for the deployment of the Internet of Things (IoT)~\cite{Asgarian2022BlueSync}.
Currently, many time-synchronization methods for WSNs use BLE~\cite{Asgarian2022BlueSync,rheinlander2016precise}.
Especially in the fields of ubiquitous computing, sports science, and health applications, BLE is most commonly used since it is integrated in modern smartphones~\cite{rheinlander2016precise}.

With BLE, timestamping errors of less than 50\,ns are possible as demonstrated by \citeauthor{Asgarian2022BlueSync} with \textit{BlueSync}~\cite{Asgarian2022BlueSync}.
\textit{BlueSync} outperforms previous work by several magnitudes \cite{rheinlander2016precise,sridhar2015cheepsync,bideaux_synchronization_2015,Ghoshdastider2014wireless,Ghoshdastider2015experimental,somaratne2018accuracy}.
While \citeauthor{rheinlander2016precise} achieve an accuracy of 0.9\,$\mu$s~\cite{rheinlander2016precise}, \citeauthor{sridhar2015cheepsync} and \citeauthor{bideaux_synchronization_2015} achieve accuracies of 10\,$\mu$s~\cite{sridhar2015cheepsync} and 14.9\,$\mu$s~\cite{bideaux_synchronization_2015}, respectively.
\citeauthor{somaratne2018accuracy}, \citeauthor{Ghoshdastider2014wireless} and \citeauthor{Ghoshdastider2015experimental} achieve accuracies of 19\,ms~\cite{somaratne2018accuracy}, 37.8\,ms~\cite{Ghoshdastider2014wireless}, and 45.1\,ms~\cite{Ghoshdastider2015experimental}, respectively.
While the above works do not have low power consumption as their main goal but achieving a low synchronization error, \citeauthor{makara_power_2019} aim also to reduce power consumption and achieve a 10\,ms error at a power consumption of 133\,$\mu$W when using two devices~\cite{makara_power_2019}.

However, there are also alternatives to BLE, including the work by \citeauthor{li_methodology_2015}, \citeauthor{Seijo2020Enhanced}, and \citeauthor{ikram2010towards} utilizing GPS~\cite{li_methodology_2015}, Wifi~\cite{Seijo2020Enhanced}, and radio signals~\cite{ikram2010towards}, respectively.
\citeauthor{li_methodology_2015} achieved synchronization with an accuracy well below 1\,ms~\cite{li_methodology_2015}, while \citeauthor{Seijo2020Enhanced} show that their method achieves sub-nanosecond time transfer accuracy~\cite{Seijo2020Enhanced}.

\subsection{Data-based Synchronization}

Multiple works have introduced data-based synchronization for distributed sensor systems using acceleration data from IMUs. 
Most of these works focus either on synchronization between IMUs and motion-capturing systems/ video data from cameras or  between only IMUs. 
The latter is typically done by subjecting all IMUs to a sudden acceleration such as shaking multiple IMUs that are held together~\cite{witchel2018thighderived, rietveld2019wheelchair, gao2019stroke, paraschiakos2020elderly, shabani_automatic_2022}. With this method, \citeauthor{shabani_automatic_2022} obtained a minimum mean lag time bias of 25.56\,ms~\cite{shabani_automatic_2022}. 
Recently, \citeauthor{spilz_novel_2021} introduced a novel data-based synchronization method for synchronizing multiple devices by subjecting them synchronously to a magnetic field impulse generated by an array of inductors causing an event that is detectable by the devices' IMUs~\cite{spilz_novel_2021} and allowing for synchronization with a maximum offset of below 2.6\,ms.
To synchronize IMUs and motion capturing systems/ video data sudden movements such as a heelstroke are commonly used. 
The event is then visible both in the video and the acceleration data~\cite{bannach_automatic_2009, plotz2012autosynch, sunwook2013motioncapture, bourke2017activitydataset, zhang2020videoacc}. 
For video-to-IMU synchronization without explicit user action, \citeauthor{zhang2020syncwise} proposed SyncWISE, an algorithm that achieves a mean error of less than half a second~\cite{zhang2020syncwise}.
When aligning recordings based on data and without distinct events, a measure for signal synchrony is required.
\citeauthor{gashi2019using} evaluated several such measures when they explored the synchrony of signals of electrodermal activity between presenters and audience members to determine engagement~\cite{gashi2019using}.

\subsection{Applications of Wearable Devices}

Activity tracking and health monitoring are among the main use cases for wearable devices.
\citeauthor{leutheuser2013hierarchical} combine data from four accelerometers in different body locations for human activity recognition~\cite{leutheuser2013hierarchical}.
At the intersection of health monitoring and activity recognition, \citeauthor{strain2020wearablesactivityhealth} perform mortality prediction based on data collected from a wrist-worn wearable~\cite{strain2020wearablesactivityhealth}.
Similarly, \citeauthor{bari2020automated} combined synchronized data from separate devices recording audio, physiological, and inertial signals to detect stressful conversations~\cite{bari2020automated}.

\section{Problem Definition and Background on Signal Alignment}

We now discuss preliminary considerations for our proposed method for synchronizing data series as well as the challenges involved.

\subsection{Temporal Alignment of Discrete Finite Signals}
\begin{figure}
    \centering
    \includegraphics[width=0.5\textwidth]{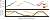}
    \caption{Example of aligning a short signal (top, orange) with a longer one (top, gray) based on cross-correlation which results in an objective function (bottom, orange) that is at its maximum where the longer signal is maximal. Similarly, the cross-covariance results in an alignment with the section of the longer signal that is the steepest (red). For 'slow' signals such as ambient pressure, alignment based on sample-by-sample signal delta is better suited (green).}
    \label{fig:xcor_prob}
\end{figure}
The problem we are addressing is specified as follows:
Given are discrete sensor recordings from two independent devices $D_1$ and $D_2$ which were acquired with temporal overlap but may be of varying duration.
A recording always comprises traces of barometer, accelerometer, and optionally other sensors.
The goal is to temporally align the two recordings based on the similarities in measurements contained in the sensor traces.
We utilize a mixture of methods based on cross-correlation, cross-covariance, and signal delta.

\subsubsection{Cross-correlation and Cross-covariance to Align Discrete Finite Signals}
The approximation of the cross-correlation $R_{fg}$ of two signals $f$ and $g$ of finite lengths $N_{f}$ and $N_{g}$ can be used to determine the best alignment of the two functions by finding the lag $m_{0}$ of $g$ that maximizes $R_{fg}$:

\[R_{fg}[m] = \frac{1}{l[m]} \sum_{n=max(0, -m)}^{min(N_{f}-1, N_{g}-m-1)} f[n]g[n+m].\]

Here, the result is scaled relative to the length $l[m]$ of the overlap of the two signals:

\[l[m] = min(N_{f}, N_{g}-m)-max(0, -m).\]

The quality of this approximation decreases at the edges with shorter window lengths $l$ when $m$ is close to $N_{g}$ or $-N_{f}$.
The cross-correlation is commonly used to align data series such as gyroscope measurements~\cite{shabani_automatic_2022} or accelerometer measurements~\cite{dargie2009analysis}, but it does not work well for all signal types.
Most notably, signals with a characteristic timescale that is longer than the duration of the computation window (i.e., contain few distinct swings within the computation window) are prone to misalignment~\cite{vio_limits_2001}.
An example of misalignment in such a scenario is shown in \autoref{fig:xcor_prob}.

In some cases, this issue can be resolved by subtracting the mean values $\mu_{f^{\prime}}$ and $\mu_{g^{\prime}}$ from the respective signals prior to multiplication. The resulting formula is an approximation of the cross-covariance of the two signals:

\[R_{fg}[m] = \frac{1}{l[m]} \sum_{n=max(0, -m)}^{min(N_{f}-1, N_{g}-m-1)} (f[n]-\mu_{f})(g[n+m]-\mu_{g})\]

However, there are still scenarios in the realm of slow sensing modalities that are misaligned when using this formula for trace alignment as shown in \autoref{fig:xcor_prob}.

\subsubsection{Signal Delta to Align Discrete Finite Signals}
\label{subsec:signal_delta_align}

To align sensor signals with a long characteristic timescale in scenarios where the sensors have recorded the same underlying signal at the same time bar the impact of sensing tolerance (as is the case for air pressure measurements), the point-by-point difference of the signals can be analyzed.
Considering the same two signals $f$ and $g$ of finite lengths $N_{f}$ and $N_{g}$ and the overlapping length $l[m]$ as before, a loss function, say $L$, can be formulated as follows:
\[L[m] = \frac{1}{l[m]} \sum_{n=max(0, -m)}^{min(N_{f}-1, N_{g}-m-1)} L_\delta(f[n] - g[n+m]) \]

Where $L_\delta$ is the Huber loss function which reduces the sensitivity to measured outliers:
  \[
     L_\delta(x)=\left\{\begin{array}{ll} \frac{x^2}{2}, & |x| \leq \delta \\
         \delta (|x| - \frac{\delta}{2}), & |x| > \delta\end{array}\right.
  \]
The resulting loss function $L[m]$ can be minimized to find the lag $m_0$ that aligns the two sensor traces the best.
Alternatively, the standard deviation of the point-by-point difference of the two signals can be used as a loss function.
This makes it invariant to the mean of the signal difference and effectively yields a measure of how well two signals run in parallel:
\[L[m] = \sqrt{\frac{1}{l[m]-1} \sum_{n=max(0, -m)}^{min(N_{f}-1, N_{g}-m-1)} \vert f[n] - g[n+m] - \mu\vert^2} \]
where $\mu$ is the mean of $f[n] - g[n+m]$:
\[\mu = \frac{1}{l[m]} \sum_{n=max(0, -m)}^{min(N_{f}-1, N_{g}-m-1)} f[n] - g[n+m] \]

\subsection{Clock Offset, Drift, and Skew} 
The difference between two RTCs $\delta(t)$ can be characterized in three components~\cite{tirado-andres_performance_2019}: (1) offset, the constant component of $\delta(t)$ caused by different starting points, (2) skew, the first derivative $\delta'(t)$ which indicates that one clock is running faster than the other resulting in a gradual de-synchronization, and (3) drift, the second derivative $\delta''(t)$ which is the variation in skew.
This terminology is consistent with previous definitions as used by \citeauthor{mills1992network}~\cite{mills1992network} and \citeauthor{sundararaman_clock_2005}~\cite{sundararaman_clock_2005}.
For RTCs that rely on quartz crystals, clock drift is mainly caused by the temperature dependence of the latter which is typically characterized as 
$Err(T) = 0.036ppm\cdot(T-25^\circ C)$~\cite{tirado-andres_performance_2019}.
This results in only minor clock drift for temperature differences commonly experienced by two wearable devices.
Therefore, we assume clock drift to be 0 and de-synchronization effects to be linear between devices in our considerations, similar to related work~\cite{sichitiu_simple_2003}.

\subsection{Measurement Errors of Air Pressure Sensors}

Air pressure measurements are subject to sensor tolerances.
The resulting error can be characterized by an absolute accuracy (i.e., the absolute difference to ground truth) and a relative accuracy (i.e., the accuracy of the delta between two measurements by the same sensor).
For commercially available barometric pressure sensors, the absolute accuracy is usually about an order of magnitude larger than the relative accuracy~\cite{bme280_datasheet, bmp581_datasheet, spl07_datasheet}.
Therefore, BMAR uses absolute measurements to reject non-simultaneous measurements only and relies on relative measurements for trace alignment.

\subsection{Ensuring within-device Sensor Synchronization}
In offline settings, the RTC of a device does not provide absolute time information which could be used for cross-device synchronization but instead keeps track of the runtime since starting the device more accurately than internal sensor clocks would.
With typical tolerances in the order of several percent, sensor clocks are not suitable to be used for timekeeping.
Hence, a prerequisite for the alignment of multiple sensor traces across devices is the synchronization of traces with the RTC within each device.

\subsubsection{Alignment Offsets Due to Sensor Startup Latency} 
Starting each sensor of one device at a known RTC timestamp does not guarantee synchronization.
This is due to differences between the latency of each sensor for activation and starting the acquisition of its first sample. 
Because this latency is not commonly specified in the datasheet, it can usually not be compensated for without extensive calibration measurements. 
However, assuming that this latency is constant for a specific sensor type and only data from the same sensor type is compared across devices, the latencies cancel out and have no negative impact on the resulting synchronization results.

\subsubsection{Synchronized Sampling} 
When sensors support external triggering, the RTC can be used to directly initiate measurements.
This keeps sensor data traces perfectly synchronized with the RTC over arbitrarily long periods of runtime.
In systems where sensors do not support external triggers from the RTC, sensors must rely on their internal clocks.
In this case, the relative error in the sampling rate requires compensation.

\section{BMAR: Method design for offline signal synchronization}

Optimal data synchronization across multiple wearable devices worn by a single person requires patterns in the data that are unique over time and do not depend on the body location where the device is worn.
The presence of such patterns depends on user behavior and sensing modality.

Therefore, BMAR builds on barometric pressure sensors, which observe changes in the environment that affect all worn devices to a similar extent.
This causes the individual on-body location of a sensor to manifest in a negligible amount and does not require our method to rely on specific user actions or activities.
Barometric pressure sensors are inexpensive, low-power, and are commonly built into wearable devices. %
The signal they measure is influenced by changes in the altitude above sea level (e.g., walking up or down stairs, taking elevators, or performing any other motion not in a flat plane), meteorological changes in atmospheric pressure, or changes in ambient pressure caused by the immediate environment (e.g., opening a door of a closed room or car). 

Our method aligns the data recorded by any two independent devices $D_1$ and $D_2$ or rejects the alignment in case it determines that the devices were not running at the same time or worn by the same person.
The goal of this process is to determine the lag time $t_{lag}(t)$ of $D2$ compared to $D1$.
$t_{lag}(t)$ itself is time-dependent due to RTC tolerances in $D1$ and $D2$ but is assumed to be linear.

\begin{figure}
    \centering
    \includegraphics[width=\textwidth]{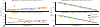}
    \caption{Approximations of the same $t_{lag}(t)$ with different $n_{win}$ for two pairs of recordings. (a) Fewer windows lead to fewer outliers and a more accurate approximation for these recordings. (b) More windows lead to a reduction in alignment error for these recordings.}
    \label{fig:ols}
\end{figure}
\subsection{Rejection of Non-simultaneous Measurements}
\label{sec:rej_traces}
The nature of air pressure measurements allows the rejection of two data traces that were not recorded at the same time. 
The difference of two temporally independent, single-sample air pressure measurements collected in the same location are expected to be normally distributed $X_1 - X_2 \sim \mathcal{N}(0,\,2\sigma^{2})$ with $\sigma^2$ being the variance of independent air pressure samples for the specific location. 
The probability of correct rejection of two such measurements can be calculated as follows:
\[p_{rej} = \frac{1}{2\sigma\sqrt{\Pi}}\int_{-2A}^{2A} e^{-\frac{x^2}{4\sigma^2}}\]
Where $\pm A$ is the range of the absolute accuracy of each barometric pressure sensor.
When inserting real-world values such as $\sigma = 928\,Pa$\,\cite{camuffo_earliest_2010} and $A=100\,Pa$\,\cite{bme280_datasheet}, the probability of correct rejection is 0.83.
Using multiple measurements over time and incorporating a distribution over altitude above sea level, this probability is expected to increase. 
Here, the sensor accuracy is assumed to have strict lower and upper bounds and thus the rate of false rejection is zero when setting the acceptance tolerance to the same boundaries.
Modeling the sensor accuracy as a normal distribution leads to the expected trade-off of minimizing false positives versus false negatives, however, specifications in datasheets usually do not provide enough information for such calculations.

To determine whether two devices $D_1$ and $D_2$ were running at the same time, we consider the alignment of the air pressure measurement data from both devices that minimizes the mean sample-by-sample difference and classify them as non-simultaneously recorded if this value surpasses a threshold.

\subsection{Stage 1: Air Pressure-based Pre-alignment}
\label{sec:pressure_prealign}
BMAR first analyzes air pressure measurement data series to determine whether $D_1$ and $D_2$ were running at the same time and to find an approximation $t^P_{lag}$ of $t_{lag}(t)$. 

Since the error of $t^P_{lag}$ is generally larger than the RTC clock skew effects between $D_1$ and $D_2$, it is computed as a constant value independent of time.
We developed two algorithms to compute air pressure-based trace alignment.
The \textit{delta-error} method computes the mean Huber loss based on the point-by-point difference between two traces for all possible alignments (Section~\ref{subsec:signal_delta_align}). 
The lag time corresponding to the alignment with the minimum mean Huber loss is selected as $t^P_{lag}$.
For this method, the data from both devices is low-pass filtered to remove noise and normalized to make the alignment invariant to constant offsets due to absolute sensor accuracy tolerance.

The \textit{delta-std} method computes the standard deviation of the point-by-point difference between two traces for all possible alignments and corresponding lag times. 
The lag time resulting in the minimum standard deviation is selected as $t^P_{lag}$.
The raw data is used for this method.

\subsection{Stage 2: Accelerometer-based Refinement and Optimization}
\label{sec:acc_refinement}

Next, BMAR leverages accelerometer signals to refine $t^P_{lag}$ and optimize the alignment.
We maximize the cross-correlation of the magnitudes to find the best alignment $t^{Accel}_{lag}(t)$.
To enable the time-dependent approximation of $t_{lag}(t)$ and to make the algorithm more robust to time-limited measurements that facilitate misalignment, the available data trace is split into $n_{win}$ windows for each of which a lag time is calculated separately.
The result of this is a series of $n_{win}$ calculated lag times temporally associated with the center point of the respective window that the original trace was split into.
We perform ordinary least squares regression (OLS) on these lag times and remove outliers using the Bonferroni outlier test.
This results in a linear approximation $t^{Accel}_{lag}(t)$. 

The optimal $n_{win}$ depends on the data.
Therefore, we repeat the computation, starting with $n_{win}=4$ and increase the number of windows by a factor of $2$ until the window size drops below $1$\,second or $n_{win}>512$.
The result of this is a set of linear approximations of $t_{lag}(t)$.
We use the $r^2$ adjusted for residual degrees of freedom of the OLS to select the most promising approximation.
This is based on the assumption that local misalignments within some of the windows do not appear linear over the whole trace, making $r^2$ an indicator of alignment quality.
\autoref{fig:ols}a  shows an example where the linear approximation with a smaller $n_{win}$ is more accurate while \autoref{fig:ols}b shows an example of an alignment where a larger $n_{win}$ leads to a better result.

\section{System implementation of BMAR}
\label{sec:implementation}

\begin{figure}
    \centering
    \includegraphics{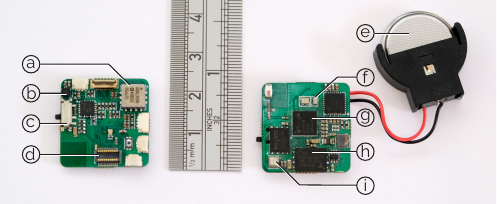}
    \caption{Top and bottom sides of the electronics of our custom wearable device, which measures 23\,mm $\times$ 22\,mm. Components: (a) ADXL355, high-resolution accelerometer, (b) LIS2DH, low-cost accelerometer, (c) power switch, (d) extension connector, used for synchronization ground truth, (e) CR2032 battery, (f) RTC quartz crystal, (g) DA14695 System-on-a-Chip, (h) TH58CYG3S0HRAIJ 8\,Gb flash memory, (i) BME280 barometric pressure sensor.}
    \label{fig:device}
\end{figure}

To implement and evaluate BMAR, we devised a miniaturized wearable platform that embeds multiple sensors and runs on a small battery.
As shown in \autoref{fig:device}, our platform is coin-sized and affords wear at multiple locations of the user's body.

\subsection{Device}

The low-power sensor device we developed comprises a 23\,mm $\times$ 22\,mm printed circuit board (PCB), which is powered by a CR2032 coin-cell battery (\autoref{fig:device}e). 
It centers around a System-on-a-Chip (SoC) (DA14695, Renesas~\cite{da14695_datasheet}, \autoref{fig:device}g) and logs sensor data to an 8\,Gb serial NAND flash memory (TH58CYG3S0HRAIJ, Kioxia America~\cite{flash_datasheet}, \autoref{fig:device}h).
The data can later be downloaded to a USB port on a PC through a serial port connection on the extension connector of the PCB (\autoref{fig:device}d). 
The RTC of the SoC relies on a 32.768kHz external quartz crystal (ECS-.327-7-16-C-TR, ECS Inc. International~\cite{rtc_quartz_datasheet}, \autoref{fig:device}f) with a tolerance of $\pm 20$\,ppm.
Our platform accommodates a series of sensors that are common on wearable devices for the purpose of evaluating our method.
Depending on the activated sensors and their sampling frequencies, the device consumes between 2 and 4.5\,mW resulting in a battery life of 35--80\,hours.
The device runs autonomously and does not require user interaction or wireless communication for operation.
It can be turned on and off by a power switch on the board (\autoref{fig:device}c).

\subsubsection{Sensors.}
For later pre-aligning of all data series, the device houses a digital barometric pressure sensor (BME280, Bosch Sensortec, \autoref{fig:device}i).
The sensor is operated in forced mode to enable synchronization to the RTC of the SoC, which triggers a measurement every 100\,ms, leading to a sampling rate of the barometer of 10\,Hz.
We used no oversampling or internal IIR filter.
As per the datasheet, the absolute accuracy of the measurements lies within a range of $\pm 1.0$\,hPa ($\pm3\sigma$) and the relative accuracy is $\pm 0.12$\,hPa ($\pm3\sigma$)~\cite{bme280_datasheet} which are low-end characteristics compared to pressure sensors currently available commercially.

For refining synchronization during our optimization stage, we included two accelerometers for comparison.
First is an ADXL355 (Analog Devices, \autoref{fig:device}a), a high-resolution 3-axis digital accelerometer.
It operates in FIFO mode, receiving triggers for measurements from the RTC of the SoC.
This results in a sampling frequency of 128\,Hz.
We set the range to $\pm2$\,g at a resolution of $20$\,bit ($4$\,$\mu$g/LSB)~\cite{adxl355_datasheet}.
Second is a LIS2DH (STMicroelectronics, \autoref{fig:device}b), a low-cost 3-axis digital accelerometer that is found in lower-end wearables such as fitness bands.
It also operates in FIFO mode, but does not support external triggers.
The accelerometer was configured to sample at 200\,Hz using the internal clock source, at a range of $\pm2$\,g and a resolution of $8$\,bit ($4$\,mg/LSB)~\cite{lis2dh_datasheet}.

\subsection{Encasing}

We created a 3D-printed case that comprises an insert to house all the electronics and the battery as well as an outer casing with a strap for wear as shown in \autoref{fig:dev_locations}b.
The case supports different straps to support wear at different locations on the body.

\subsection{Synchronization}

The firmware of the SoC was implemented in C++, building on freeRTOS to drive all sensors and repeatedly download and store the contents of their FIFO buffers.
We implemented all synchronization algorithms in Python 3.11 to process all recordings offline after download.

\subsubsection{Buffered Sampling With Compensation} 
Since the low-cost accelerometer does not support external measurement triggering from the RTC but relies on its internal clock source, its measurements require compensation to preserve the synchronization of all sensors within the device. 
When the SoC retrieves buffers from a sensor, it additionally logs an RTC-based timestamp to memory.
During offline processing, the expected total data length after each FIFO readout can be calculated based on the logged RTC timestamps of the readout and the sensor activation.
To account for the latency between sensor activation and data availability in the FIFO, we calculate the expected length relative to the first FIFO readout:
\[l_{exp}(t_n) = (t_n-t_0)\cdot f_{sensor} + l(t_0)\]
With the sampling frequency of the sensor $f_{sensor}$, the expected length $l_{exp}$ at the RTC timestamp $t_n$ of the $n^{th}$ FIFO readout after the initial readout at RTC timestamp $t_0$ after which a total of $l(t_0)$ samples were present. 
Because FIFO readouts are usually not performed on a perfectly regular basis, a lack of samples after a FIFO read-out at $t_n$ (i.e., $l(t_n)<l_{exp}(t_n)$) may automatically be compensated for by a sample which narrowly slipped into the next FIFO readout. 
We account for this by only adding or removing samples to match the expected length when no inverse operation is necessary in close temporal proximity.
This results in a sensor data trace which is synchronized to the device's RTC with a tolerance of $1/f_{sensor}$.

\subsubsection{Hyperparameters}
For the rejection of non-simultaneous traces (Section~\ref{sec:rej_traces}), we set the threshold of the mean  delta between two traces to the absolute tolerance of the used barometric pressure sensor of $100\,Pa$\,\cite{bme280_datasheet}, trying to achieve a recall close to 1. 
Lowering it increases precision and lowers recall.
We use the same value for the $\delta$ of the Huber loss in the delta-error method (Section~\ref{sec:pressure_prealign}) such that it only affects rarely occurring outliers in the measurements.
We set the maximum range of the accelerometer-based refinement (Section~\ref{sec:acc_refinement}) to $\pm 5$\,seconds. 
If the pre-alignment differs more than that, the refinement cannot fully compensate for it. 
On the other hand, increasing this range increases the risk of misalignment caused by this stage.

\section{Evaluation}

To quantify the performance of BMAR, we conducted a data recording experiment during everyday activities, using between 2 and 5 devices at once with several repetitions.
\autoref{tab:eval_overview} shows an overview of the configurations in our evaluation.
Since there is no impact of individual users' behavior on our method, all recordings were conducted by one experimenter.

\begin{table}[t]

    \caption{For our evaluation, the experimenter recorded data during these activities, wearing multiple devices at the respective locations on the body for the given durations. Each activity was repeated two or more times.}
    \centering
    \begin{tabular}{lllll}
          activity & \# & locations & runtime & rep \\
        \midrule
         sleep & 5 & head, chest, ankle, wrist & 9.9 h & 4 \\
         work & 3 & head, chest, ankle & 3.8 h & 3 \\
         cycle & 4-5 & head, chest, ankle, wrist & 1.4 h & 2 \\
         ski & 3 & chest, wrist & 8.8 h & 2
    \end{tabular}
    \label{tab:eval_overview}
\end{table}

To analyze the performance of our method, we consider pairs of devices that either did or did not record data at the same time while worn by the experimenter.
This means that if 3 devices were worn at the same time, we considered them as 3 pairs of simultaneously running devices in our evaluation ($D_1$ and $D_2$, $D_1$ and $D_3$, $D_2$ and $D_3$) and also included pairings of each device with a random device from another recording which was not acquired at the same time.

\subsection{Procedure}
\begin{figure}[b]
    \centering
    \includegraphics{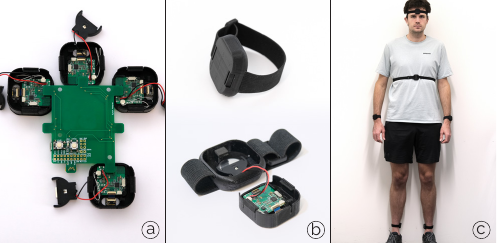}
    \caption{Apparatus implementation for the evaluation.
    (a)~Board to connect multiple devices and trigger ground-truth synchronization events,
    (b)~3D printed case for the custom wearable sensing devices, here shown with a wrist strap,
    (c)~the 6 sensor locations used in the evaluation of BMAR.}
    \label{fig:dev_locations}
\end{figure}

\subsubsection{Locations.}

As shown in \autoref{fig:dev_locations}c, devices were attached to the following locations for a recording session: strapped onto the forehead, strapped onto the sternum, on one of the wrists like a watch, or above one of the lateral malleoli.

\subsubsection{Activities.}

The experiment included two categories of sessions: active wearer and passive wearer.
For active sessions, the experimenter performed activities such as cycling, skiing, or office work over the full duration.
For this session type, we collected 31 pairs of device recordings running at the same time, with an average runtime of 3.73\,hours.

For passive sessions, the experimenter remained stationary over the full duration. We recorded these data during sleep, turning on the devices after getting into bed and turning them off before exiting.
For this session type, we obtained 40 pairs of devices with simultaneous recordings and an average runtime of 9.9\,hours.
Due to the lack of activities and motion, this session type can be considered a worst-case baseline for our analysis.

\subsection{Ground-truth Synchronization}

We devised an apparatus to establish repeated ground-truth synchronization across all devices.
The experimenter manually attached this apparatus to all devices at the beginning and the end of each session.
This resulted in two ground-truth synchronization events in our logs for reference.
The apparatus is shown in \autoref{fig:dev_locations}a.
It sends a trigger from a manual, debounced push-button to a digital pin of the SoC in each device.
The tethered connection ensured accurate reference values without any impact of wireless communication.

To ensure that each device could register the precise moment of a button press, we used a timer block on the SoC to support more finely-grained timing resolution between the RTC ticks. 
The incoming synchronization signal was logged through the timers event capture functionality, which logs the exact timestamp and triggers an interrupt to process the event. 
In theory, the accuracy of this mechanism is limited by our timer clock of 32.768\,kHz, which corresponds to a tolerance of 0.03\,ms. 
However, the tolerance of the timer event trigger capture is not specified by the SoC manufacturer.
Hardware limitations (e.g., stray capacitances and tolerances in logic voltage level detection) may also increase this error.

\subsubsection{Accuracy of Ground-truth Synchronization.}

To evaluate the accuracy in practice, we sent multiple synchronization signals to all connected devices in short succession within a few seconds. 
Assuming no clock skew between the first and last synchronization signal, we compared the alignment of the multiple detected events on the time axis.
Across 349 such synchronization sequences with 3 signal triggers each, we found a mean error of $\text{ME}=1.2\,\text{ms}$ ($\sigma=1.1\,\text{ms}$) for a single signal trigger.
To improve accuracy in the evaluation, we always used the mean across 3 consecutive signal triggers.

\begin{figure}[b]
    \centering
    \includegraphics[width=0.6\textwidth]{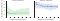}
    \caption{Median alignment error of BMAR's synchronization. The band denotes the range between 10th and 90th percentile.}
    \label{fig:overall_accuracy}
\end{figure}

\subsection{Duration of Recordings}
\label{subsec:duration_of_recordings}

For temporal alignment evaluations, the duration of the recording has an impact on the results (i.e., the probability of correct temporal alignment increases with longer recordings).
We conducted the evaluations considering various different lengths of recordings by cropping recordings to the respective length.
If a recording was multiple times the desired length, we split it and used up to 5 non-overlapping splits separately.
The number of traces for different lengths is shown in \autoref{tab:error}.
When aligning traces, we only cropped or split one of the two traces and preserved the length of the other trace.
We considered the length of the shorter trace to be the minimum overlapping time of the two traces to prevent negative effects on the edges.
Effectively, this means we measured the accuracy of aligning a trace of a specific length to a longer one that overlaps it.

\begin{figure}[b]
    \centering
    \includegraphics[width=0.6\textwidth]{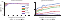}
    \caption{Percentage of traces that BMAR correctly aligned with a given error tolerance.}
    \label{fig:overall_percentiles}
\end{figure}

\subsection{Results}
We first report the results of our complete method spanning both air pressure-based pre-alignment and accelerometer-based refinement.
We then report the results of the rejection of non-simultaneously recorded data, of pre-alignment, and of refinement separately.

\begin{table}[t]
    \centering
    
    \caption{Median alignment error in seconds of the complete synchronization method. The number of pairs of recordings used is stated in brackets.}
    \begin{tabular}{lllllll}
    \multirow{2}{*}{\shortstack[l]{}}
        &\multicolumn{6}{c}{length of recording}\\
    
        & 5\,min & 10\,min & 15\,min & 30\,min & 1\,h & $\ge6$\,h \\
        \midrule
        active& 0.442 (155)& 0.074 (155)& 0.044 (155)& 0.041 (107)& 0.028 (73)& 0.034 (31)\\
        passive&$\gg$60 (200)&$\gg$60 (200)&$\gg$60 (200)&20.2 (200)&5.42 (200)&3.06 (40)\\

    \end{tabular}
    \label{tab:error}
\end{table}

\subsubsection{Overall Synchronization Accuracy}
\label{subsec:overall_results}
As shown in \autoref{fig:overall_accuracy} and \autoref{tab:error}, the synchronization of active sessions achieved a median alignment error of 33.4\,ms for recording durations of 15\,min and longer, 73.6\,ms for durations of 10\,min, and 443\,ms for durations of 5\,min.
The synchronization of passive sessions (i.e., a worst-case baseline without user motion or activity) resulted in a median alignment error of 3.06\,s for recording durations of 6\,h and longer, 5.42\,s for durations of 2\,h, and 20.2\,s for durations of 30\,min.
\autoref{fig:overall_percentiles} shows the percentage of synchronizations that our method aligned within a given error tolerance.

We report the errors of unconstrained synchronizations with median and percentiles because the error of alignment for outliers (i.e., failed, effectively random alignments) can be up to four orders of magnitude larger than the median error of all alignments.
The impact on the mean error and standard deviation makes them less informative for evaluation.

\begin{figure}[t]
    \centering
    \includegraphics[width=0.56\textwidth]{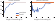}
    \caption{Precision and recall rates for BMAR's detection of simultaneously acquired recordings.}
    \label{fig:simult_class}
\end{figure}

\begin{figure}[b]
    \centering
    \includegraphics[width=0.5\textwidth]{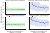}
    \vspace{-2mm}
    \caption{Median alignment error of BMAR's first stage (air pressure-based pre-alignment).
    The band denotes the range between the 10th and 90th percentile.}
    \label{fig:p_prealign}
    \vspace{-2mm}
\end{figure}

\subsubsection{Accuracy of Stage\,1: Pressure-based Pre-alignment}
\label{subsec:p_results}
In stage 1, air pressure measurments were used to detect non-simultaneous measurements and do a first trace alignment of simultaneously acquired traces.

\paragraph{Elimination of non-simultaneous measurements.}
When using a sensor tolerance of 100\,Pa as the threshold for the maximum mean difference between two traces to detect simultaneously recorded air pressure data, we reach a precision and recall of 1.0 for recordings of active sessions that contain at least 35\,minutes of data and recordings of inactive sessions of at least 8\,hours of recording time.
\autoref{fig:simult_class} shows the precision and recall rates for recordings of different lengths.

These results are based on tests with the barometric pressure sensor recordings of each device $D_n$ both paired with simultaneously and randomly selected, non-simultaneously recorded data.
There is a trade-off between precision and recall:
Lowering the threshold improves precision but lowers recall.
However, such a lowered recall can generally not be compensated for by increasing the duration of the recording.

\begin{figure}[t]
  \centering
  \subfloat[high-resolution accelerometer]{
    \includegraphics[width=0.48\textwidth]{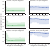}
    \label{fig:imu_refin_high_res}
  }
  \hfill
  \subfloat[standard-resolution accelerometer]{
    \includegraphics[width=0.48\textwidth]{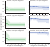}
    \label{fig:imu_refin_standard_res}
  }
  \caption{Median alignment error after BMAR's second stage (accelerometer-based refinement) for different pre-alignment errors.}
\end{figure}

\paragraph{Trace alignment based on air pressure.}

Aligning active sessions produced a median alignment error of 1.14\,s for recording durations of 10\,min and longer, 1.53\,s for durations of 5\,min, and 3.8\,s for durations of 3\,min, as shown in \autoref{fig:p_prealign}.
The synchronization of passive sessions (i.e., our worst-case baseline) resulted in a median alignment error of 3.36\,s for recording durations of 3\,h and longer, 12.4\,s for durations of 1\,h, and 25.3\,s for durations of 30\,min.

Of the two tested alignment methods, the \textit{delta-std} method performed better than the \textit{delta-error} method.
Across all recording lengths greater than 0.5\,hours (i.e., after initial convergence), the median alignment error of the  \textit{delta-std} is on average 48\,ms lower for active sessions and 237\,ms lower for passive sessions.

\subsubsection{Accuracy of Stage\,2: Accelerometer-based Refinement}
\label{subsec:refinement}

Testing the accelerometer-based synchronization refinement in isolation for active sessions and under the assumption of a pre-alignment tolerance of $\pm 1\,s$, produced a median alignment error of 32.0\,ms for recording durations of 15\,min and longer, 40.6\,ms for durations of 5\,min, and 218\,ms for durations of 1\,min, as shown in Figure~\ref{fig:imu_refin_high_res}.
The refinement of passive sessions (i.e., worst-case baseline) resulted in a median alignment error of 274\,ms for recording durations of 15\,min and longer, 580\,ms for durations of 5\,min, and 606\,ms for durations of 1\,min. As shown in Figure~\ref{fig:imu_refin_high_res}, the median error increases with higher pre-alignment tolerances.

We tested the refinement with data from both a high-resolution accelerometer (ADXL355, Analog Devices) and a standard-resolution accelerometer (LIS2DH, STMicroelectronics).
The high-resolution accelerometer produced better results, yielding an average alignment error that was 21.5\,ms lower when analyzing all recording durations, which resulted in a median alignment error below 0.1\,s.

As shown in Figure~\ref{fig:imu_refin_standard_res}, the results we obtained with the standard-resolution accelerometer for active sessions and under the assumption of a pre-alignment tolerance of $\pm 1\,s$, produced a median alignment error of 44.2\,ms for recording durations of 15\,min and longer, 71.9\,ms for durations of 5\,min, and 276\,ms for durations of 1\,min.

The refinement of passive sessions (i.e., worst-case baseline) based on the standard-resolution accelerometer resulted in a median alignment error of 345\,ms for recording durations of 15\,min and longer, 685\,ms for durations of 5\,min, and 703\,ms for durations of 1\,min.

\section{Discussion, Limitations, and Future Work}

Our evaluation confirmed the robustness of our method to detect recordings that were acquired simultaneously to synchronize data stemming from multiple wearable devices.
The accuracy our method achieved for recordings during active sessions is high enough to correctly match each occurrence of fast repetitive events across devices.
This includes the fastest vital sign (i.e., heartbeats of a child at 3.5 \,Hz) and fast body motions (i.e., the cadence of a professional sprinter at 4 \,Hz or the catch rate of the world's fastest juggler at 9\,Hz).
Faster motions by humans usually occur with longer intervals in between, making cross-device analysis easier.
Unsurprisingly, the results were worse for our baseline recordings, which did not include any activity other than sleeping.
However, given long recordings of multiple hours, non-simultaneous recordings are still rejected reliably and the synchronization tolerance reaches 3.06\,seconds which may be sufficient for the analysis of changes in sleep position.

\subsection{The Relevance of Pre-alignment Using Air Pressure Measurements}

Due to differences in the motion characteristics across body parts, IMU data does not lend itself well to aligning data traces without additional constraints.
Using cross-correlation, for example, the resulting objective function exhibits many maxima and trace alignment without restrictive constraints on the search window or specifically performed motions by the wearer is challenging.
For a more general, rough alignment, the use of a sensing modality with a longer characteristic timescale is more effective to produce a cost function that monotonically converges to a global minimum.

Temperature, perhaps the most commonly logged sensing modality, does include features common across devices worn in different body locations.
Events like moving from indoors to outdoors with an abrupt change in ambient temperature could be used for synchronization, but they may be attenuated or delayed due to varying thermal insulation of the devices by clothing.

Air pressure is thus a suitable modality for cross-device synchronization: 
Air pressure measurements are barely affected by anything other than the geographical location, time of recording, and sensor tolerance.
The former two are identical for multiple devices worn simultaneously.

\subsection{Impact of Sensor Accuracy}

As mentioned in Section~\ref{subsec:refinement}, our standard-resolution accelerometer performed similarly well as the high-resolution accelerometer during refinement, indicating that our method does not require an expensive accelerometer for accurate signal synchronization.
The range of commercially available barometric pressure sensors is smaller compared to accelerometers and the sensor used for the evaluation can be considered to be at the low end of the spectrum in terms of accuracy:
In the relevant temperature and measurement range, the BME280 we evaluated BMAR with has an absolute and relative accuracy of $\pm100Pa$ and $\pm12Pa$, respectively~\cite{bme280_datasheet}.
This compares to $\pm30Pa$ and $\pm6Pa$ for the recently released BMP581 (Bosch Sensortec)~\cite{bmp581_datasheet} as well as $\pm50Pa$ and $\pm3Pa$ for the inexpensive SPL07-003 (Goertek Microelectronics Inc.)~\cite{spl07_datasheet}.
This suggests that our results likely generalize to most available pressure sensors, including inexpensive commodity barometric pressure sensors.

We expect a better absolute accuracy to improve the rejection of non-simultaneous recordings and a better relative accuracy to improve the pre-alignment of the recordings.
In the future, we plan to characterize the specific impact of sensor characteristics (accuracy, sampling rate, resolution, noise, and RTC synchronization) on the synchronization error. 

\subsection{Needed Overlap of Cross-sensor Signal Recordings}

In our current design, BMAR aligns signals of a specified duration $l_1$ to a recording of length $l_2$, which fully contains the former (Section~\ref{subsec:duration_of_recordings}).
However, our method also works in scenarios where signals only \emph{partially} overlap.
In these, the effectively smaller overlap decreases the amount of data for calculating synchronization cost and may lead to a higher risk of misalignment.
Considering a minimum overlap of $l'\leq l_1$, the expected error is in the range of our evaluation results for total recording lengths between $l_1$ and $l'$.

\subsection{Limitations of Accelerometer-based Refinements}

The improvement in refining synchronization based on accelerometer signals depends on the signal characteristics.
For our method, we cannot give worst-case guarantees;
if recordings contain no events or asynchronous patterns, our refinement may worsen the result of the pre-alignment.

To prevent potentially negative effects of our refinement, BMAR limits the range within which pre-alignment sequences may be refined as part of our synchronization method.
Since the loss of the accelerometer-based alignment does not generally decrease monotonically to a correct global minimum, the refinement may worsen the result of a pre-alignment that was successful but with an error larger than the range of the refinement.%
\autoref{fig:imu_refin_high_res} shows how our refinement for short recordings during inactive sessions is close to a random offset within alignment tolerance and only decreases for longer recordings.
Future work could improve our approach by leveraging the pre-alignment confidence and suitability of the accelerometer data to dynamically determine the range of the refinement.

\subsection{Devices across Multiple Users and Locations}

BMAR is currently designed to synchronize signals captured on devices worn by a single person.
However, our method of aligning signal traces based on air pressure recordings could extend to synchronizing signal streams captured across multiple people and beyond the realm of wearable devices.
We expect the method to produce accurate results if all devices are in the same location or move collectively (i.e., in a car, train, or elevator), but we anticipate a negative impact of movements that are spread out between devices, such as when groups of people who wear them do not transition between building floors at the same time.%
Depending on the setting, the use of different sensing modalities may be appropriate such as sound or temperature.
We tested the latter early on in our project but found that for wearable devices, the impact of clothing and body temperature makes cross-device synchronization challenging.
It may, however, be a viable option when synchronizing non-wearable devices.

\subsection{Limitations of BMAR's Evaluation: Single Participant vs. Breadth of Activities for More Insight}

We acknowledge that in our evaluation, multiple devices collected data on a single participant only.
While such an evaluation may be uncommon, it is important to point out that BMAR is not specific to human factors or capabilities; rather, the performance of the method depends on the type of activities.%
Therefore, the breadth of assessed activities and environments is considerably more important in evaluating our method than the number of participants, each of which would perform activities only slightly differently.

In our evaluation, the sensor devices were affixed to five locations on the participant's body, without a strict enforcement of inter-device distance, which covered a range of device constellations and variability similar to varying body heights across participants.
Our results showed no notable impact of the body location on the accuracy of trace alignment.

Most importantly, our evaluation has shown that the most challenging activity for our cross-sensor synchronization approach is no activity (i.e., resting and sleeping). 
While we believe that our results generalize and an evaluation with multiple participants would not change the findings, we acknowledge that for assessing this conclusively, our evaluation could be extended to multiple participants in the future.

\subsection{Speculative Comparison with Online Synchronization}
Online synchronization through a wireless network may be a viable alternative to BMAR to reach higher accuracy (Section~\ref{subsec:rw_wireless_sync}).
A main advantage of BMAR is its low power consumption: At the sampling rates used in our evaluation, a barometric pressure sensor operates on less than $35\,\mu W$~\cite{bmp581_datasheet} and the low-resolution accelerometer uses less than $70\,\mu W$~\cite{lis2dh_datasheet}.
Of the commonly used wireless protocols, currently, only Bluetooth low energy reaches comparable values: \citeauthor{makara_power_2019} synchronized two nodes with a power consumption of 133\,uW with a typical accuracy of 10\,ms.

However, producing reliable implementations in practice is challenging, limiting the choice of SoC, and no prior work has shown synchronization of more than two nodes with power consumption in this order of magnitude.
In such a scenario, a more powerful device such as a smartphone may be needed as central device for the other nodes to remain in the stated range of power consumption while BMARs power requirements are unaffected by the number of devices.

Generally, most existing synchronization algorithms are optimized for low error---not low power consumption, which has thus often not been evaluated in prior work.
Because most wearable devices already routinely sample acceleration and/or air pressure from built-in sensors, our method imposes no additional power consumption during runtime.
Additionally, our method may be used with devices that are not engineered to be compatible to be used in the same wireless network.

\section{Conclusion}

We have presented BMAR, a novel method to synchronize sensor data across multiple wearable devices without the need for communication or explicit user interaction during runtime.
BMAR leverages barometric pressure sensors and accelerometers, which are inexpensive, low-power, and commonly found in wearable devices.
Therefore, they represent a simple solution to synchronizing wearable devices without the need for inter-device communication, allowing data recording and device operation to focus on low power consumption in miniaturized devices.

BMAR's two-stage process comprises pre-alignment, which we use to avoid latching on to local but not a global optimum in alignment, and refinement.
We evaluated BMAR using custom-built, low-power wearable sensing devices that recorded a series of signals across a variety of activities over the course of up to 10\,hours.
Our method achieved accurate synchronization with a median error of 33.4\,ms, which is low enough for most practical purposes.
Even during activities with little to no motion such as sleeping, our method synchronized signals with an error of 3\,s, all while reliably rejecting signals that were not concurrently recorded across devices. 

We believe that BMAR is a practical approach for supporting cross-device data collections that build on unobtrusive miniaturized devices, focus on duration, and ease of use.

\bibliographystyle{ACM-Reference-Format}
\bibliography{references, references_2, references_max, datasheets}

%%% -*-BibTeX-*-
%%% Do NOT edit. File created by BibTeX with style
%%% ACM-Reference-Format-Journals [18-Jan-2012].

\begin{thebibliography}{62}

%%% ====================================================================
%%% NOTE TO THE USER: you can override these defaults by providing
%%% customized versions of any of these macros before the \bibliography
%%% command.  Each of them MUST provide its own final punctuation,
%%% except for \shownote{}, \showDOI{}, and \showURL{}.  The latter two
%%% do not use final punctuation, in order to avoid confusing it with
%%% the Web address.
%%%
%%% To suppress output of a particular field, define its macro to expand
%%% to an empty string, or better, \unskip, like this:
%%%
%%% \newcommand{\showDOI}[1]{\unskip}   % LaTeX syntax
%%%
%%% \def \showDOI #1{\unskip}           % plain TeX syntax
%%%
%%% ====================================================================

\ifx \showCODEN    \undefined \def \showCODEN     #1{\unskip}     \fi
\ifx \showDOI      \undefined \def \showDOI       #1{#1}\fi
\ifx \showISBNx    \undefined \def \showISBNx     #1{\unskip}     \fi
\ifx \showISBNxiii \undefined \def \showISBNxiii  #1{\unskip}     \fi
\ifx \showISSN     \undefined \def \showISSN      #1{\unskip}     \fi
\ifx \showLCCN     \undefined \def \showLCCN      #1{\unskip}     \fi
\ifx \shownote     \undefined \def \shownote      #1{#1}          \fi
\ifx \showarticletitle \undefined \def \showarticletitle #1{#1}   \fi
\ifx \showURL      \undefined \def \showURL       {\relax}        \fi
% The following commands are used for tagged output and should be
% invisible to TeX
\providecommand\bibfield[2]{#2}
\providecommand\bibinfo[2]{#2}
\providecommand\natexlab[1]{#1}
\providecommand\showeprint[2][]{arXiv:#2}

\bibitem[Alinia et~al\mbox{.}(2017)]%
        {Alinia2017wearables}
\bibfield{author}{\bibinfo{person}{Parastoo Alinia}, \bibinfo{person}{Chris
  Cain}, \bibinfo{person}{Ramin Fallahzadeh}, \bibinfo{person}{Armin
  Shahrokni}, \bibinfo{person}{Diane Cook}, {and} \bibinfo{person}{Hassan
  Ghasemzadeh}.} \bibinfo{year}{2017}\natexlab{}.
\newblock \showarticletitle{How Accurate Is Your Activity Tracker? A
  Comparative Study of Step Counts in {Low-Intensity} Physical Activities}.
\newblock \bibinfo{journal}{\emph{JMIR Mhealth Uhealth}} \bibinfo{volume}{5},
  \bibinfo{number}{8} (\bibinfo{date}{Aug.} \bibinfo{year}{2017}),
  \bibinfo{pages}{e106}.
\newblock


\bibitem[America(2019)]%
        {flash_datasheet}
\bibfield{author}{\bibinfo{person}{Kioxia America}.}
  \bibinfo{year}{2019}\natexlab{}.
\newblock \bibinfo{title}{TH58CYG3S0HRAIJ}.
\newblock
  \bibinfo{howpublished}{\url{https://static6.arrow.com/aropdfconversion/dd02224cfed190f11368fa73736fc9a9b13d6528/th58cyg3s0hraij_datasheet_en_20191001.pdf}}.
\newblock
\newblock
\shownote{Last accessed: 2022-12-09}.


\bibitem[An et~al\mbox{.}(2017)]%
        {an2017wearable}
\bibfield{author}{\bibinfo{person}{Byeong~Wan An}, \bibinfo{person}{Jung~Hwal
  Shin}, \bibinfo{person}{So-Yun Kim}, \bibinfo{person}{Joohee Kim},
  \bibinfo{person}{Sangyoon Ji}, \bibinfo{person}{Jihun Park},
  \bibinfo{person}{Youngjin Lee}, \bibinfo{person}{Jiuk Jang},
  \bibinfo{person}{Young-Geun Park}, \bibinfo{person}{Eunjin Cho},
  \bibinfo{person}{Subin Jo}, {and} \bibinfo{person}{Jang-Ung Park}.}
  \bibinfo{year}{2017}\natexlab{}.
\newblock \showarticletitle{Smart Sensor Systems for Wearable Electronic
  Devices}.
\newblock \bibinfo{journal}{\emph{Polymers}} \bibinfo{volume}{9},
  \bibinfo{number}{8} (\bibinfo{year}{2017}).
\newblock
\showISSN{2073-4360}
\urldef\tempurl%
\url{https://doi.org/10.3390/polym9080303}
\showDOI{\tempurl}


\bibitem[Asgarian and Najafi(2022)]%
        {Asgarian2022BlueSync}
\bibfield{author}{\bibinfo{person}{Farzad Asgarian} {and}
  \bibinfo{person}{Khalil Najafi}.} \bibinfo{year}{2022}\natexlab{}.
\newblock \showarticletitle{BlueSync: Time Synchronization in Bluetooth Low
  Energy With Energy-Efficient Calculations}.
\newblock \bibinfo{journal}{\emph{IEEE Internet of Things Journal}}
  \bibinfo{volume}{9}, \bibinfo{number}{11} (\bibinfo{year}{2022}),
  \bibinfo{pages}{8633--8645}.
\newblock
\urldef\tempurl%
\url{https://doi.org/10.1109/JIOT.2021.3116921}
\showDOI{\tempurl}


\bibitem[Bannach et~al\mbox{.}(2009)]%
        {bannach_automatic_2009}
\bibfield{author}{\bibinfo{person}{David Bannach}, \bibinfo{person}{Oliver
  Amft}, {and} \bibinfo{person}{Paul Lukowicz}.}
  \bibinfo{year}{2009}\natexlab{}.
\newblock \showarticletitle{Automatic {Event}-{Based} {Synchronization} of
  {Multimodal} {Data} {Streams} from {Wearable} and {Ambient} {Sensors}}. In
  \bibinfo{booktitle}{\emph{Smart {Sensing} and {Context}}}
  \emph{(\bibinfo{series}{Lecture {Notes} in {Computer} {Science}})},
  \bibfield{editor}{\bibinfo{person}{Payam Barnaghi}, \bibinfo{person}{Klaus
  Moessner}, \bibinfo{person}{Mirko Presser}, {and} \bibinfo{person}{Stefan
  Meissner}} (Eds.). \bibinfo{publisher}{Springer}, \bibinfo{address}{Berlin,
  Heidelberg}, \bibinfo{pages}{135--148}.
\newblock
\showISBNx{978-3-642-04471-7}
\urldef\tempurl%
\url{https://doi.org/10.1007/978-3-642-04471-7_11}
\showDOI{\tempurl}


\bibitem[Bari et~al\mbox{.}(2020)]%
        {bari2020automated}
\bibfield{author}{\bibinfo{person}{Rummana Bari}, \bibinfo{person}{Md~Mahbubur
  Rahman}, \bibinfo{person}{Nazir Saleheen}, \bibinfo{person}{Megan~Battles
  Parsons}, \bibinfo{person}{Eugene~H Buder}, {and} \bibinfo{person}{Santosh
  Kumar}.} \bibinfo{year}{2020}\natexlab{}.
\newblock \showarticletitle{Automated detection of stressful conversations
  using wearable physiological and inertial sensors}.
\newblock \bibinfo{journal}{\emph{Proceedings of the ACM on interactive,
  mobile, wearable and ubiquitous technologies}} \bibinfo{volume}{4},
  \bibinfo{number}{4} (\bibinfo{year}{2020}), \bibinfo{pages}{1--23}.
\newblock


\bibitem[Bideaux et~al\mbox{.}(2015)]%
        {bideaux_synchronization_2015}
\bibfield{author}{\bibinfo{person}{André Bideaux}, \bibinfo{person}{Bernd
  Zimmermann}, \bibinfo{person}{Stefan Hey}, {and} \bibinfo{person}{Wilhelm
  Stork}.} \bibinfo{year}{2015}\natexlab{}.
\newblock \showarticletitle{Synchronization in wireless biomedical-sensor
  networks with {Bluetooth} {Low} {Energy}}.
\newblock \bibinfo{journal}{\emph{Current Directions in Biomedical
  Engineering}} \bibinfo{volume}{1}, \bibinfo{number}{1} (\bibinfo{date}{Sept.}
  \bibinfo{year}{2015}), \bibinfo{pages}{73--76}.
\newblock
\urldef\tempurl%
\url{https://doi.org/10.1515/cdbme-2015-0019}
\showDOI{\tempurl}
\newblock
\shownote{Publisher: De Gruyter Section: Current Directions in Biomedical
  Engineering}.


\bibitem[Bisht et~al\mbox{.}(2021)]%
        {bisht_study_2021}
\bibfield{author}{\bibinfo{person}{Rajendra~Singh Bisht},
  \bibinfo{person}{Sourabh Jain}, {and} \bibinfo{person}{Naveen Tewari}.}
  \bibinfo{year}{2021}\natexlab{}.
\newblock \showarticletitle{Study of {Wearable} {IoT} devices in 2021:
  {Analysis} \& {Future} {Prospects}}. In \bibinfo{booktitle}{\emph{2021 2nd
  {International} {Conference} on {Intelligent} {Engineering} and {Management}
  ({ICIEM})}}. \bibinfo{pages}{577--581}.
\newblock
\urldef\tempurl%
\url{https://doi.org/10.1109/ICIEM51511.2021.9445334}
\showDOI{\tempurl}


\bibitem[Bloss(2015)]%
        {Bloss2015wearable}
\bibfield{author}{\bibinfo{person}{Richard Bloss}.}
  \bibinfo{year}{2015}\natexlab{}.
\newblock \showarticletitle{Wearable sensors bring new benefits to continuous
  medical monitoring, real time physical activity assessment, baby monitoring
  and industrial applications}.
\newblock \bibinfo{journal}{\emph{Sensor Review}} \bibinfo{volume}{35},
  \bibinfo{number}{2} (\bibinfo{date}{Jan.} \bibinfo{year}{2015}),
  \bibinfo{pages}{141--145}.
\newblock


\bibitem[Bourke et~al\mbox{.}(2017)]%
        {bourke2017activitydataset}
\bibfield{author}{\bibinfo{person}{Alan~Kevin Bourke}, \bibinfo{person}{Espen
  Alexander~F Ihlen}, \bibinfo{person}{Ronny Bergquist},
  \bibinfo{person}{Per~Bendik Wik}, \bibinfo{person}{Beatrix Vereijken}, {and}
  \bibinfo{person}{Jorunn~L Helbostad}.} \bibinfo{year}{2017}\natexlab{}.
\newblock \showarticletitle{A Physical Activity Reference {Data-Set} Recorded
  from Older Adults Using {Body-Worn} Inertial Sensors and Video
  {Technology-The} {ADAPT} Study {Data-Set}}.
\newblock \bibinfo{journal}{\emph{Sensors (Basel)}} \bibinfo{volume}{17},
  \bibinfo{number}{3} (\bibinfo{date}{March} \bibinfo{year}{2017}).
\newblock


\bibitem[Bruno et~al\mbox{.}(2021)]%
        {bruno2021battery}
\bibfield{author}{\bibinfo{person}{Elisa Bruno}, \bibinfo{person}{Sebastian
  Böttcher}, \bibinfo{person}{Pedro~F. Viana}, \bibinfo{person}{Marta
  Amengual-Gual}, \bibinfo{person}{Boney Joseph}, \bibinfo{person}{Nino
  Epitashvili}, \bibinfo{person}{Matthias Dümpelmann}, \bibinfo{person}{Martin
  Glasstetter}, \bibinfo{person}{Andrea Biondi}, \bibinfo{person}{Kristof
  Van~Laerhoven}, \bibinfo{person}{Tobias Loddenkemper},
  \bibinfo{person}{Mark~P. Richardson}, \bibinfo{person}{Andreas
  Schulze-Bonhage}, {and} \bibinfo{person}{Benjamin~H. Brinkmann}.}
  \bibinfo{year}{2021}\natexlab{}.
\newblock \showarticletitle{Wearable devices for seizure detection: Practical
  experiences and recommendations from the Wearables for Epilepsy And Research
  (WEAR) International Study Group}.
\newblock \bibinfo{journal}{\emph{Epilepsia}} \bibinfo{volume}{62},
  \bibinfo{number}{10} (\bibinfo{year}{2021}), \bibinfo{pages}{2307--2321}.
\newblock
\urldef\tempurl%
\url{https://doi.org/10.1111/epi.17044}
\showDOI{\tempurl}
\showeprint{https://onlinelibrary.wiley.com/doi/pdf/10.1111/epi.17044}


\bibitem[Camuffo et~al\mbox{.}(2010)]%
        {camuffo_earliest_2010}
\bibfield{author}{\bibinfo{person}{Dario Camuffo}, \bibinfo{person}{Chiara
  Bertolin}, \bibinfo{person}{Phil~D. Jones}, \bibinfo{person}{Richard Cornes},
  {and} \bibinfo{person}{Emmanuel Garnier}.} \bibinfo{year}{2010}\natexlab{}.
\newblock \showarticletitle{The earliest daily barometric pressure readings in
  {Italy}: {Pisa} {AD} 1657-1658 and {Modena} {AD} 1694, and the weather over
  {Europe}}.
\newblock \bibinfo{journal}{\emph{The Holocene}} \bibinfo{volume}{20},
  \bibinfo{number}{3} (\bibinfo{date}{May} \bibinfo{year}{2010}),
  \bibinfo{pages}{337--349}.
\newblock
\showISSN{0959-6836}
\urldef\tempurl%
\url{https://doi.org/10.1177/0959683609351900}
\showDOI{\tempurl}
\newblock
\shownote{Publisher: SAGE Publications Ltd}.


\bibitem[Case et~al\mbox{.}(2015)]%
        {Case2015wearable}
\bibfield{author}{\bibinfo{person}{Meredith~A Case}, \bibinfo{person}{Holland~A
  Burwick}, \bibinfo{person}{Kevin~G Volpp}, {and} \bibinfo{person}{Mitesh~S
  Patel}.} \bibinfo{year}{2015}\natexlab{}.
\newblock \showarticletitle{Accuracy of Smartphone Applications and Wearable
  Devices for Tracking Physical Activity Data}.
\newblock \bibinfo{journal}{\emph{JAMA}} \bibinfo{volume}{313},
  \bibinfo{number}{6} (\bibinfo{date}{Feb.} \bibinfo{year}{2015}),
  \bibinfo{pages}{625--626}.
\newblock


\bibitem[Choi and Shin(2017)]%
        {choi_photoplethysmography_2017}
\bibfield{author}{\bibinfo{person}{A. Choi} {and} \bibinfo{person}{H. Shin}.}
  \bibinfo{year}{2017}\natexlab{}.
\newblock \showarticletitle{Photoplethysmography sampling frequency: pilot
  assessment of how low can we go to analyze pulse rate variability with
  reliability?}
\newblock \bibinfo{journal}{\emph{Physiological Measurement}}
  \bibinfo{volume}{38}, \bibinfo{number}{3} (\bibinfo{date}{Feb.}
  \bibinfo{year}{2017}), \bibinfo{pages}{586}.
\newblock
\showISSN{0967-3334}
\urldef\tempurl%
\url{https://doi.org/10.1088/1361-6579/aa5efa}
\showDOI{\tempurl}
\newblock
\shownote{Publisher: IOP Publishing}.


\bibitem[Dahmen et~al\mbox{.}(2017)]%
        {dahmen2017wearables}
\bibfield{author}{\bibinfo{person}{Jordana Dahmen}, \bibinfo{person}{Alyssa
  La~Fleur}, \bibinfo{person}{Gina Sprint}, \bibinfo{person}{Diane Cook}, {and}
  \bibinfo{person}{Douglas~L. Weeks}.} \bibinfo{year}{2017}\natexlab{}.
\newblock \showarticletitle{Using wrist-worn sensors to measure and compare
  physical activity changes for patients undergoing rehabilitation}. In
  \bibinfo{booktitle}{\emph{2017 IEEE International Conference on Pervasive
  Computing and Communications Workshops (PerCom Workshops)}}.
  \bibinfo{pages}{667--672}.
\newblock
\urldef\tempurl%
\url{https://doi.org/10.1109/PERCOMW.2017.7917643}
\showDOI{\tempurl}


\bibitem[Dana and Penrod(1990)]%
        {dana1990role}
\bibfield{author}{\bibinfo{person}{Peter~H Dana} {and} \bibinfo{person}{Bruce~M
  Penrod}.} \bibinfo{year}{1990}\natexlab{}.
\newblock \showarticletitle{The role of GPS in precise time and frequency
  dissemination}.
\newblock \bibinfo{journal}{\emph{GPS World}} \bibinfo{volume}{1},
  \bibinfo{number}{4} (\bibinfo{year}{1990}), \bibinfo{pages}{38--43}.
\newblock


\bibitem[Dargie(2009)]%
        {dargie2009analysis}
\bibfield{author}{\bibinfo{person}{Waltenegus Dargie}.}
  \bibinfo{year}{2009}\natexlab{}.
\newblock \showarticletitle{Analysis of time and frequency domain features of
  accelerometer measurements}. In \bibinfo{booktitle}{\emph{2009 Proceedings of
  18th International Conference on Computer Communications and Networks}}.
  IEEE, \bibinfo{pages}{1--6}.
\newblock


\bibitem[Devices(2018)]%
        {adxl355_datasheet}
\bibfield{author}{\bibinfo{person}{Analog Devices}.}
  \bibinfo{year}{2018}\natexlab{}.
\newblock \bibinfo{title}{ADXL355, Low Noise, Low Drift, Low Power, 3-Axis MEMS
  Accelerometers}.
\newblock
  \bibinfo{howpublished}{\url{https://www.analog.com/media/en/technical-documentation/data-sheets/adxl354_355.pdf}}.
\newblock
\newblock
\shownote{Last accessed: 2022-12-09}.


\bibitem[Electronics(2022)]%
        {da14695_datasheet}
\bibfield{author}{\bibinfo{person}{Renesas Electronics}.}
  \bibinfo{year}{2022}\natexlab{}.
\newblock \bibinfo{title}{DA1469x}.
\newblock
  \bibinfo{howpublished}{\url{https://www.renesas.com/us/en/document/dst/da1469x-datasheet?r=1606281}}.
\newblock
\newblock
\shownote{Last accessed: 2022-12-09}.


\bibitem[Elson et~al\mbox{.}(2002)]%
        {elson_fine-grained_2002}
\bibfield{author}{\bibinfo{person}{Jeremy Elson}, \bibinfo{person}{Lewis
  Girod}, {and} \bibinfo{person}{Deborah Estrin}.}
  \bibinfo{year}{2002}\natexlab{}.
\newblock \showarticletitle{Fine-{Grained} {Network} {Time} {Synchronization}
  {Using} {Reference} {Broadcasts}}.
\newblock In \bibinfo{booktitle}{\emph{5th {Symposium} on {Operating} {Systems}
  {Design} and {Implementation} ({OSDI} 02)}}. \bibinfo{publisher}{USENIX
  Association}.
\newblock
\urldef\tempurl%
\url{https://www.usenix.org/conference/osdi-02/fine-grained-network-time-synchronization-using-reference-broadcasts}
\showURL{%
\tempurl}


\bibitem[Ganeriwal et~al\mbox{.}(2003)]%
        {ganeriwal_timing-sync_2003}
\bibfield{author}{\bibinfo{person}{Saurabh Ganeriwal}, \bibinfo{person}{Ram
  Kumar}, {and} \bibinfo{person}{Mani~B. Srivastava}.}
  \bibinfo{year}{2003}\natexlab{}.
\newblock \showarticletitle{Timing-sync protocol for sensor networks}. In
  \bibinfo{booktitle}{\emph{Proceedings of the 1st international conference on
  {Embedded} networked sensor systems}} \emph{(\bibinfo{series}{{SenSys}
  '03})}. \bibinfo{publisher}{Association for Computing Machinery},
  \bibinfo{address}{New York, NY, USA}, \bibinfo{pages}{138--149}.
\newblock
\showISBNx{978-1-58113-707-1}
\urldef\tempurl%
\url{https://doi.org/10.1145/958491.958508}
\showDOI{\tempurl}


\bibitem[Gao et~al\mbox{.}(2019)]%
        {gao2019stroke}
\bibfield{author}{\bibinfo{person}{Yan Gao}, \bibinfo{person}{Yang Long},
  \bibinfo{person}{Yu Guan}, \bibinfo{person}{Anna Basu},
  \bibinfo{person}{Jessica Baggaley}, {and} \bibinfo{person}{Thomas Ploetz}.}
  \bibinfo{year}{2019}\natexlab{}.
\newblock \showarticletitle{Towards Reliable, Automated General Movement
  Assessment for Perinatal Stroke Screening in Infants Using Wearable
  Accelerometers}.
\newblock \bibinfo{journal}{\emph{Proc. ACM Interact. Mob. Wearable Ubiquitous
  Technol.}} \bibinfo{volume}{3}, \bibinfo{number}{1}, Article
  \bibinfo{articleno}{12} (\bibinfo{date}{mar} \bibinfo{year}{2019}),
  \bibinfo{numpages}{22}~pages.
\newblock
\urldef\tempurl%
\url{https://doi.org/10.1145/3314399}
\showDOI{\tempurl}


\bibitem[Gashi et~al\mbox{.}(2019)]%
        {gashi2019using}
\bibfield{author}{\bibinfo{person}{Shkurta Gashi}, \bibinfo{person}{Elena
  Di~Lascio}, {and} \bibinfo{person}{Silvia Santini}.}
  \bibinfo{year}{2019}\natexlab{}.
\newblock \showarticletitle{Using unobtrusive wearable sensors to measure the
  physiological synchrony between presenters and audience members}.
\newblock \bibinfo{journal}{\emph{Proceedings of the ACM on Interactive,
  Mobile, Wearable and Ubiquitous Technologies}} \bibinfo{volume}{3},
  \bibinfo{number}{1} (\bibinfo{year}{2019}), \bibinfo{pages}{1--19}.
\newblock


\bibitem[Ghoshdastider et~al\mbox{.}(2014)]%
        {Ghoshdastider2014wireless}
\bibfield{author}{\bibinfo{person}{Unmesh Ghoshdastider},
  \bibinfo{person}{Reinhard Viga}, {and} \bibinfo{person}{Michael Kraft}.}
  \bibinfo{year}{2014}\natexlab{}.
\newblock \showarticletitle{Wireless time synchronization of a collaborative
  brain-computer-interface using bluetooth low energy}. In
  \bibinfo{booktitle}{\emph{SENSORS, 2014 IEEE}}. \bibinfo{pages}{2250--2254}.
\newblock
\urldef\tempurl%
\url{https://doi.org/10.1109/ICSENS.2014.6985489}
\showDOI{\tempurl}


\bibitem[Ghoshdastider et~al\mbox{.}(2015)]%
        {Ghoshdastider2015experimental}
\bibfield{author}{\bibinfo{person}{Unmesh Ghoshdastider},
  \bibinfo{person}{Reinhard Viga}, {and} \bibinfo{person}{Michael Kraft}.}
  \bibinfo{year}{2015}\natexlab{}.
\newblock \showarticletitle{Experimental evaluation of a pairwise broadcast
  synchronization in a low-power Cyber-physical system}. In
  \bibinfo{booktitle}{\emph{2015 IEEE Topical Conference on Wireless Sensors
  and Sensor Networks (WiSNet)}}. \bibinfo{pages}{50--52}.
\newblock
\urldef\tempurl%
\url{https://doi.org/10.1109/WISNET.2015.7127399}
\showDOI{\tempurl}


\bibitem[Gilbert et~al\mbox{.}(2022)]%
        {gilbert_simple_2022}
\bibfield{author}{\bibinfo{person}{Thomas Gilbert}, \bibinfo{person}{Sally
  Day}, \bibinfo{person}{Antonia F De~C Hamilton}, {and} \bibinfo{person}{Jamie
  Ward}.} \bibinfo{year}{2022}\natexlab{}.
\newblock \showarticletitle{A {Simple} {Method} for {Synchronising} {Multiple}
  {IMUs} using the {Magnetometer}}. In \bibinfo{booktitle}{\emph{Proceedings of
  the 2022 {ACM} {International} {Symposium} on {Wearable} {Computers}}}
  \emph{(\bibinfo{series}{{ISWC} '22})}. \bibinfo{publisher}{Association for
  Computing Machinery}, \bibinfo{address}{New York, NY, USA},
  \bibinfo{pages}{100--102}.
\newblock
\showISBNx{978-1-4503-9424-6}
\urldef\tempurl%
\url{https://doi.org/10.1145/3544794.3558466}
\showDOI{\tempurl}


\bibitem[Gjoreski et~al\mbox{.}(2016)]%
        {Gjoreski2016wearable}
\bibfield{author}{\bibinfo{person}{Martin Gjoreski}, \bibinfo{person}{Hristijan
  Gjoreski}, \bibinfo{person}{Mitja Lu{\v s}trek}, {and}
  \bibinfo{person}{Matja{\v z} Gams}.} \bibinfo{year}{2016}\natexlab{}.
\newblock \showarticletitle{How Accurately Can Your Wrist Device Recognize
  Daily Activities and Detect Falls?}
\newblock \bibinfo{journal}{\emph{Sensors (Basel)}} \bibinfo{volume}{16},
  \bibinfo{number}{6} (\bibinfo{date}{June} \bibinfo{year}{2016}).
\newblock


\bibitem[Ikram et~al\mbox{.}(2010)]%
        {ikram2010towards}
\bibfield{author}{\bibinfo{person}{Waqas Ikram}, \bibinfo{person}{Ivan
  Stoianov}, {and} \bibinfo{person}{Nina~F Thornhill}.}
  \bibinfo{year}{2010}\natexlab{}.
\newblock \showarticletitle{Towards a radio-controlled time synchronized
  wireless sensor network: A work in-progress paper}. In
  \bibinfo{booktitle}{\emph{2010 IEEE 15th Conference on Emerging Technologies
  \& Factory Automation (ETFA 2010)}}. IEEE, \bibinfo{pages}{1--4}.
\newblock


\bibitem[Inc.(2022)]%
        {spl07_datasheet}
\bibfield{author}{\bibinfo{person}{Goertek~Microelectronics Inc.}}
  \bibinfo{year}{2022}\natexlab{}.
\newblock \bibinfo{title}{SPL07-003, Digital pressure sensor}.
\newblock
  \bibinfo{howpublished}{\url{https://media.digikey.com/pdf/Data\%20Sheets/Goertek\%20Microelectronics\%20PDFs/SPL07-003.pdf}}.
\newblock
\newblock
\shownote{Last accessed: 2023-02-14}.


\bibitem[International(2021)]%
        {rtc_quartz_datasheet}
\bibfield{author}{\bibinfo{person}{ECS~Inc International}.}
  \bibinfo{year}{2021}\natexlab{}.
\newblock \bibinfo{title}{ECX-16, 32.768 KHz SMD Tuning Fork Crystal}.
\newblock
  \bibinfo{howpublished}{\url{https://ecsxtal.com/store/pdf/ECX-16.pdf}}.
\newblock
\newblock
\shownote{Last accessed: 2022-12-09}.


\bibitem[Kamišalić et~al\mbox{.}(2018)]%
        {kamisalic2018wearablesactivityhealth}
\bibfield{author}{\bibinfo{person}{Aida Kamišalić}, \bibinfo{person}{Iztok
  Fister}, \bibinfo{person}{Muhamed Turkanović}, {and} \bibinfo{person}{Sašo
  Karakatič}.} \bibinfo{year}{2018}\natexlab{}.
\newblock \showarticletitle{Sensors and Functionalities of Non-Invasive
  Wrist-Wearable Devices: A Review}.
\newblock \bibinfo{journal}{\emph{Sensors}} \bibinfo{volume}{18},
  \bibinfo{number}{6} (\bibinfo{year}{2018}).
\newblock
\showISSN{1424-8220}
\urldef\tempurl%
\url{https://doi.org/10.3390/s18061714}
\showDOI{\tempurl}


\bibitem[Kim and Nussbaum(2013)]%
        {sunwook2013motioncapture}
\bibfield{author}{\bibinfo{person}{Sunwook Kim} {and} \bibinfo{person}{Maury~A.
  Nussbaum}.} \bibinfo{year}{2013}\natexlab{}.
\newblock \showarticletitle{Performance evaluation of a wearable inertial
  motion capture system for capturing physical exposures during manual material
  handling tasks}.
\newblock \bibinfo{journal}{\emph{Ergonomics}} \bibinfo{volume}{56},
  \bibinfo{number}{2} (\bibinfo{year}{2013}), \bibinfo{pages}{314--326}.
\newblock
\urldef\tempurl%
\url{https://doi.org/10.1080/00140139.2012.742932}
\showDOI{\tempurl}
\showeprint{https://doi.org/10.1080/00140139.2012.742932}
\newblock
\shownote{PMID: 23231730}.


\bibitem[Leutheuser et~al\mbox{.}(2013)]%
        {leutheuser2013hierarchical}
\bibfield{author}{\bibinfo{person}{Heike Leutheuser}, \bibinfo{person}{Dominik
  Schuldhaus}, {and} \bibinfo{person}{Bjoern~M Eskofier}.}
  \bibinfo{year}{2013}\natexlab{}.
\newblock \showarticletitle{Hierarchical, multi-sensor based classification of
  daily life activities: comparison with state-of-the-art algorithms using a
  benchmark dataset}.
\newblock \bibinfo{journal}{\emph{PloS one}} \bibinfo{volume}{8},
  \bibinfo{number}{10} (\bibinfo{year}{2013}), \bibinfo{pages}{e75196}.
\newblock


\bibitem[Li et~al\mbox{.}(2015)]%
        {li_methodology_2015}
\bibfield{author}{\bibinfo{person}{Zan Li}, \bibinfo{person}{Torsten Braun},
  {and} \bibinfo{person}{Desislava~C. Dimitrova}.}
  \bibinfo{year}{2015}\natexlab{}.
\newblock \showarticletitle{Methodology for {GPS} {Synchronization}
  {Evaluation} with {High} {Accuracy}}. In \bibinfo{booktitle}{\emph{2015
  {IEEE} 81st {Vehicular} {Technology} {Conference} ({VTC} {Spring})}}.
  \bibinfo{pages}{1--6}.
\newblock
\urldef\tempurl%
\url{https://doi.org/10.1109/VTCSpring.2015.7145929}
\showDOI{\tempurl}
\newblock
\shownote{ISSN: 1550-2252}.


\bibitem[Liu et~al\mbox{.}(2016)]%
        {Liu2016wearable}
\bibfield{author}{\bibinfo{person}{Xi Liu}, \bibinfo{person}{Lei Liu},
  \bibinfo{person}{Steven~J. Simske}, {and} \bibinfo{person}{Jerry Liu}.}
  \bibinfo{year}{2016}\natexlab{}.
\newblock \showarticletitle{Human Daily Activity Recognition for Healthcare
  Using Wearable and Visual Sensing Data}. In \bibinfo{booktitle}{\emph{2016
  IEEE International Conference on Healthcare Informatics (ICHI)}}.
  \bibinfo{pages}{24--31}.
\newblock
\urldef\tempurl%
\url{https://doi.org/10.1109/ICHI.2016.100}
\showDOI{\tempurl}


\bibitem[Luz et~al\mbox{.}(2014)]%
        {luz_evaluating_2014}
\bibfield{author}{\bibinfo{person}{Eduardo José da~S. Luz},
  \bibinfo{person}{David Menotti}, {and} \bibinfo{person}{William~Robson
  Schwartz}.} \bibinfo{year}{2014}\natexlab{}.
\newblock \showarticletitle{Evaluating the use of {ECG} signal in low
  frequencies as a biometry}.
\newblock \bibinfo{journal}{\emph{Expert Systems with Applications}}
  \bibinfo{volume}{41}, \bibinfo{number}{5} (\bibinfo{date}{April}
  \bibinfo{year}{2014}), \bibinfo{pages}{2309--2315}.
\newblock
\showISSN{0957-4174}
\urldef\tempurl%
\url{https://doi.org/10.1016/j.eswa.2013.09.028}
\showDOI{\tempurl}


\bibitem[Makara et~al\mbox{.}(2019)]%
        {makara_power_2019}
\bibfield{author}{\bibinfo{person}{Dmytro Makara}, \bibinfo{person}{Vladyslav
  Tsybul’nyk}, {and} \bibinfo{person}{Taras Kurnyts’kyi}.}
  \bibinfo{year}{2019}\natexlab{}.
\newblock \showarticletitle{Power {Efficient} {Clock} {Synchronization} in
  {Bluetooth}-{Based} {Mesh} {Networks}}. In \bibinfo{booktitle}{\emph{Ambient
  {Intelligence}}} \emph{(\bibinfo{series}{Lecture {Notes} in {Computer}
  {Science}})}, \bibfield{editor}{\bibinfo{person}{Ioannis Chatzigiannakis},
  \bibinfo{person}{Boris De~Ruyter}, {and} \bibinfo{person}{Irene Mavrommati}}
  (Eds.). \bibinfo{publisher}{Springer International Publishing},
  \bibinfo{address}{Cham}, \bibinfo{pages}{14--26}.
\newblock
\showISBNx{978-3-030-34255-5}
\urldef\tempurl%
\url{https://doi.org/10.1007/978-3-030-34255-5_2}
\showDOI{\tempurl}


\bibitem[Malhi et~al\mbox{.}(2012)]%
        {malhi2012zigbee}
\bibfield{author}{\bibinfo{person}{Karandeep Malhi},
  \bibinfo{person}{Subhas~Chandra Mukhopadhyay}, \bibinfo{person}{Julia
  Schnepper}, \bibinfo{person}{Mathias Haefke}, {and} \bibinfo{person}{Hartmut
  Ewald}.} \bibinfo{year}{2012}\natexlab{}.
\newblock \showarticletitle{A Zigbee-Based Wearable Physiological Parameters
  Monitoring System}.
\newblock \bibinfo{journal}{\emph{IEEE Sensors Journal}} \bibinfo{volume}{12},
  \bibinfo{number}{3} (\bibinfo{year}{2012}), \bibinfo{pages}{423--430}.
\newblock
\urldef\tempurl%
\url{https://doi.org/10.1109/JSEN.2010.2091719}
\showDOI{\tempurl}


\bibitem[Maróti et~al\mbox{.}(2004)]%
        {maroti_flooding_2004}
\bibfield{author}{\bibinfo{person}{Miklós Maróti}, \bibinfo{person}{Branislav
  Kusy}, \bibinfo{person}{Gyula Simon}, {and} \bibinfo{person}{Ákos
  Lédeczi}.} \bibinfo{year}{2004}\natexlab{}.
\newblock \showarticletitle{The flooding time synchronization protocol}. In
  \bibinfo{booktitle}{\emph{Proceedings of the 2nd international conference on
  {Embedded} networked sensor systems}} \emph{(\bibinfo{series}{{SenSys}
  '04})}. \bibinfo{publisher}{Association for Computing Machinery},
  \bibinfo{address}{New York, NY, USA}, \bibinfo{pages}{39--49}.
\newblock
\showISBNx{978-1-58113-879-5}
\urldef\tempurl%
\url{https://doi.org/10.1145/1031495.1031501}
\showDOI{\tempurl}


\bibitem[Mills(1992)]%
        {mills1992network}
\bibfield{author}{\bibinfo{person}{David Mills}.}
  \bibinfo{year}{1992}\natexlab{}.
\newblock \bibinfo{booktitle}{\emph{Network time protocol (version 3)
  specification, implementation and analysis}}.
\newblock \bibinfo{type}{{T}echnical {R}eport}.
\newblock


\bibitem[Paraschiakos et~al\mbox{.}(2020)]%
        {paraschiakos2020elderly}
\bibfield{author}{\bibinfo{person}{Stylianos Paraschiakos},
  \bibinfo{person}{Ricardo Cachucho}, \bibinfo{person}{Matthijs Moed},
  \bibinfo{person}{Diana van Heemst}, \bibinfo{person}{Simon Mooijaart},
  \bibinfo{person}{Eline~P Slagboom}, \bibinfo{person}{Arno Knobbe}, {and}
  \bibinfo{person}{Marian Beekman}.} \bibinfo{year}{2020}\natexlab{}.
\newblock \showarticletitle{Activity recognition using wearable sensors for
  tracking the elderly}.
\newblock \bibinfo{journal}{\emph{User Modeling and User-Adapted Interaction}}
  \bibinfo{volume}{30}, \bibinfo{number}{3} (\bibinfo{date}{July}
  \bibinfo{year}{2020}), \bibinfo{pages}{567--605}.
\newblock


\bibitem[Park and Jayaraman(2017)]%
        {park_wearables_2017}
\bibfield{author}{\bibinfo{person}{Sungmee Park} {and}
  \bibinfo{person}{Sundaresan Jayaraman}.} \bibinfo{year}{2017}\natexlab{}.
\newblock \showarticletitle{The wearables revolution and {Big} {Data}: the
  textile lineage}.
\newblock \bibinfo{journal}{\emph{The Journal of The Textile Institute}}
  \bibinfo{volume}{108}, \bibinfo{number}{4} (\bibinfo{date}{April}
  \bibinfo{year}{2017}), \bibinfo{pages}{605--614}.
\newblock
\showISSN{0040-5000}
\urldef\tempurl%
\url{https://doi.org/10.1080/00405000.2016.1176632}
\showDOI{\tempurl}
\newblock
\shownote{Publisher: Taylor \& Francis \_eprint:
  https://doi.org/10.1080/00405000.2016.1176632}.


\bibitem[Plotz et~al\mbox{.}(2012)]%
        {plotz2012autosynch}
\bibfield{author}{\bibinfo{person}{Thomas Plotz}, \bibinfo{person}{Chen Chen},
  \bibinfo{person}{Nils~Y. Hammerla}, {and} \bibinfo{person}{Gregory~D.
  Abowd}.} \bibinfo{year}{2012}\natexlab{}.
\newblock \showarticletitle{Automatic Synchronization of Wearable Sensors and
  Video-Cameras for Ground Truth Annotation -- A Practical Approach}. In
  \bibinfo{booktitle}{\emph{2012 16th International Symposium on Wearable
  Computers}}. \bibinfo{pages}{100--103}.
\newblock
\urldef\tempurl%
\url{https://doi.org/10.1109/ISWC.2012.15}
\showDOI{\tempurl}


\bibitem[Rheinl{\"a}nder and Wehn(2016)]%
        {rheinlander2016precise}
\bibfield{author}{\bibinfo{person}{Carl~C Rheinl{\"a}nder} {and}
  \bibinfo{person}{Norbert Wehn}.} \bibinfo{year}{2016}\natexlab{}.
\newblock \showarticletitle{Precise synchronization time stamp generation for
  Bluetooth low energy}. In \bibinfo{booktitle}{\emph{2016 IEEE SENSORS}}.
  IEEE, \bibinfo{pages}{1--3}.
\newblock


\bibitem[Rietveld et~al\mbox{.}(2019)]%
        {rietveld2019wheelchair}
\bibfield{author}{\bibinfo{person}{Thomas Rietveld}, \bibinfo{person}{Riemer
  J.~K. Vegter}, \bibinfo{person}{Rienk M.~A. van~der Slikke},
  \bibinfo{person}{Aldo~E. Hoekstra}, \bibinfo{person}{Lucas H.~V. van~der
  Woude}, {and} \bibinfo{person}{Sonja de Groot}.}
  \bibinfo{year}{2019}\natexlab{}.
\newblock \showarticletitle{Wheelchair mobility performance of elite wheelchair
  tennis players during four field tests: Inter-trial reliability and construct
  validity}.
\newblock \bibinfo{journal}{\emph{PLOS ONE}} \bibinfo{volume}{14},
  \bibinfo{number}{6} (\bibinfo{date}{06} \bibinfo{year}{2019}),
  \bibinfo{pages}{1--16}.
\newblock
\urldef\tempurl%
\url{https://doi.org/10.1371/journal.pone.0217514}
\showDOI{\tempurl}


\bibitem[Seijo et~al\mbox{.}(2020)]%
        {Seijo2020Enhanced}
\bibfield{author}{\bibinfo{person}{Óscar Seijo},
  \bibinfo{person}{Jesús~Alberto López-Fernández},
  \bibinfo{person}{Hans-Peter Bernhard}, {and} \bibinfo{person}{Iñaki Val}.}
  \bibinfo{year}{2020}\natexlab{}.
\newblock \showarticletitle{Enhanced Timestamping Method for Subnanosecond Time
  Synchronization in IEEE 802.11 Over WLAN Standard Conditions}.
\newblock \bibinfo{journal}{\emph{IEEE Transactions on Industrial Informatics}}
  \bibinfo{volume}{16}, \bibinfo{number}{9} (\bibinfo{year}{2020}),
  \bibinfo{pages}{5792--5805}.
\newblock
\urldef\tempurl%
\url{https://doi.org/10.1109/TII.2019.2959200}
\showDOI{\tempurl}


\bibitem[Sensortec(2022a)]%
        {bme280_datasheet}
\bibfield{author}{\bibinfo{person}{Bosch Sensortec}.}
  \bibinfo{year}{2022}\natexlab{a}.
\newblock \bibinfo{title}{BME280, Combined humidity and pressure sensor}.
\newblock
  \bibinfo{howpublished}{\url{https://www.bosch-sensortec.com/media/boschsensortec/downloads/datasheets/bst-bme280-ds002.pdf}}.
\newblock
\newblock
\shownote{Last accessed: 2022-12-09}.


\bibitem[Sensortec(2022b)]%
        {bmp581_datasheet}
\bibfield{author}{\bibinfo{person}{Bosch Sensortec}.}
  \bibinfo{year}{2022}\natexlab{b}.
\newblock \bibinfo{title}{BMP581, Barometric Pressure Sensor}.
\newblock
  \bibinfo{howpublished}{\url{https://www.bosch-sensortec.com/media/boschsensortec/downloads/datasheets/bst-bmp581-ds004.pdf}}.
\newblock
\newblock
\shownote{Last accessed: 2022-12-09}.


\bibitem[Shabani et~al\mbox{.}(2022)]%
        {shabani_automatic_2022}
\bibfield{author}{\bibinfo{person}{Shaban Shabani}, \bibinfo{person}{Alan~K.
  Bourke}, \bibinfo{person}{Amir Muaremi}, \bibinfo{person}{Jens Praestgaard},
  \bibinfo{person}{Kate O'Keeffe}, \bibinfo{person}{Rob Argent},
  \bibinfo{person}{Martin Brom}, \bibinfo{person}{Celeste Scotti},
  \bibinfo{person}{Brian Caulfield}, {and} \bibinfo{person}{Lorcan~C. Walsh}.}
  \bibinfo{year}{2022}\natexlab{}.
\newblock \showarticletitle{An {Automatic} {Foot} and {Shank} {IMU}
  {Synchronization} {Algorithm}: {Proof}-of-concept}. In
  \bibinfo{booktitle}{\emph{2022 44th {Annual} {International} {Conference} of
  the {IEEE} {Engineering} in {Medicine} \& {Biology} {Society} ({EMBC})}}.
  \bibinfo{pages}{4210--4213}.
\newblock
\urldef\tempurl%
\url{https://doi.org/10.1109/EMBC48229.2022.9871162}
\showDOI{\tempurl}
\newblock
\shownote{ISSN: 2694-0604}.


\bibitem[Sichitiu and Veerarittiphan(2003)]%
        {sichitiu_simple_2003}
\bibfield{author}{\bibinfo{person}{M.L. Sichitiu} {and} \bibinfo{person}{C.
  Veerarittiphan}.} \bibinfo{year}{2003}\natexlab{}.
\newblock \showarticletitle{Simple, accurate time synchronization for wireless
  sensor networks}. In \bibinfo{booktitle}{\emph{2003 {IEEE} {Wireless}
  {Communications} and {Networking}, 2003. {WCNC} 2003.}},
  Vol.~\bibinfo{volume}{2}. \bibinfo{pages}{1266--1273 vol.2}.
\newblock
\urldef\tempurl%
\url{https://doi.org/10.1109/WCNC.2003.1200555}
\showDOI{\tempurl}
\newblock
\shownote{ISSN: 1525-3511}.


\bibitem[Somaratne et~al\mbox{.}(2018)]%
        {somaratne2018accuracy}
\bibfield{author}{\bibinfo{person}{Kasun Somaratne}, \bibinfo{person}{F~John
  Dian}, {and} \bibinfo{person}{Amirhossein Yousefi}.}
  \bibinfo{year}{2018}\natexlab{}.
\newblock \showarticletitle{Accuracy analysis of time synchronization using
  current consumption pattern of BLE devices}. In
  \bibinfo{booktitle}{\emph{2018 IEEE 8th Annual Computing and Communication
  Workshop and Conference (CCWC)}}. IEEE, \bibinfo{pages}{841--844}.
\newblock


\bibitem[Spilz and Munz(2021)]%
        {spilz_novel_2021}
\bibfield{author}{\bibinfo{person}{Andreas Spilz} {and}
  \bibinfo{person}{Michael Munz}.} \bibinfo{year}{2021}\natexlab{}.
\newblock \bibinfo{title}{Novel {Approach} {To} {Synchronisation} {Of}
  {Wearable} {IMUs} {Based} {On} {Magnetometers}}.
\newblock
\newblock
\urldef\tempurl%
\url{https://doi.org/10.48550/arXiv.2107.03147}
\showDOI{\tempurl}
\newblock
\shownote{arXiv:2107.03147 [cs, eess]}.


\bibitem[Sridhar et~al\mbox{.}(2015)]%
        {sridhar2015cheepsync}
\bibfield{author}{\bibinfo{person}{Sabarish Sridhar}, \bibinfo{person}{Prasant
  Misra}, {and} \bibinfo{person}{Jay Warrior}.}
  \bibinfo{year}{2015}\natexlab{}.
\newblock \showarticletitle{CheepSync: {A} Time Synchronization Service for
  Resource Constrained Bluetooth Low Energy Advertisers}.
\newblock \bibinfo{journal}{\emph{CoRR}}  \bibinfo{volume}{abs/1501.06479}
  (\bibinfo{year}{2015}).
\newblock
\showeprint[arXiv]{1501.06479}
\urldef\tempurl%
\url{http://arxiv.org/abs/1501.06479}
\showURL{%
\tempurl}


\bibitem[STMicroelectronics(2011)]%
        {lis2dh_datasheet}
\bibfield{author}{\bibinfo{person}{STMicroelectronics}.}
  \bibinfo{year}{2011}\natexlab{}.
\newblock \bibinfo{title}{LIS2DH, MEMS digital output motion sensor: ultra
  low-power high performance 3-axis “femto” accelerometer}.
\newblock
  \bibinfo{howpublished}{\url{https://www.st.com/resource/en/datasheet/lis2dh.pdf}}.
\newblock
\newblock
\shownote{Last accessed: 2022-12-09}.


\bibitem[Strain et~al\mbox{.}(2020)]%
        {strain2020wearablesactivityhealth}
\bibfield{author}{\bibinfo{person}{Tessa Strain}, \bibinfo{person}{Katrien
  Wijndaele}, \bibinfo{person}{Paddy~C Dempsey}, \bibinfo{person}{Stephen~J
  Sharp}, \bibinfo{person}{Matthew Pearce}, \bibinfo{person}{Justin Jeon},
  \bibinfo{person}{Tim Lindsay}, \bibinfo{person}{Nick Wareham}, {and}
  \bibinfo{person}{S{\o}ren Brage}.} \bibinfo{year}{2020}\natexlab{}.
\newblock \showarticletitle{Wearable-device-measured physical activity and
  future health risk}.
\newblock \bibinfo{journal}{\emph{Nature Medicine}} \bibinfo{volume}{26},
  \bibinfo{number}{9} (\bibinfo{date}{Sept.} \bibinfo{year}{2020}),
  \bibinfo{pages}{1385--1391}.
\newblock


\bibitem[Sundararaman et~al\mbox{.}(2005)]%
        {sundararaman_clock_2005}
\bibfield{author}{\bibinfo{person}{Bharath Sundararaman}, \bibinfo{person}{Ugo
  Buy}, {and} \bibinfo{person}{Ajay~D. Kshemkalyani}.}
  \bibinfo{year}{2005}\natexlab{}.
\newblock \showarticletitle{Clock synchronization for wireless sensor networks:
  a survey}.
\newblock \bibinfo{journal}{\emph{Ad Hoc Networks}} \bibinfo{volume}{3},
  \bibinfo{number}{3} (\bibinfo{date}{May} \bibinfo{year}{2005}),
  \bibinfo{pages}{281--323}.
\newblock
\showISSN{1570-8705}
\urldef\tempurl%
\url{https://doi.org/10.1016/j.adhoc.2005.01.002}
\showDOI{\tempurl}


\bibitem[Tirado-Andrés and Araujo(2019)]%
        {tirado-andres_performance_2019}
\bibfield{author}{\bibinfo{person}{Francisco Tirado-Andrés} {and}
  \bibinfo{person}{Alvaro Araujo}.} \bibinfo{year}{2019}\natexlab{}.
\newblock \showarticletitle{Performance of clock sources and their influence on
  time synchronization in wireless sensor networks}.
\newblock \bibinfo{journal}{\emph{International Journal of Distributed Sensor
  Networks}}  \bibinfo{volume}{15} (\bibinfo{date}{Sept.}
  \bibinfo{year}{2019}), \bibinfo{pages}{155014771987937}.
\newblock
\urldef\tempurl%
\url{https://doi.org/10.1177/1550147719879372}
\showDOI{\tempurl}


\bibitem[Vio and Wamsteker(2001)]%
        {vio_limits_2001}
\bibfield{author}{\bibinfo{person}{Roberto Vio} {and} \bibinfo{person}{Willem
  Wamsteker}.} \bibinfo{year}{2001}\natexlab{}.
\newblock \showarticletitle{Limits of the {Cross}‐{Correlation} {Function} in
  the {Analysis} of {Short} {Time} {Series}}.
\newblock \bibinfo{journal}{\emph{Publications of the Astronomical Society of
  the Pacific}} \bibinfo{volume}{113}, \bibinfo{number}{779}
  (\bibinfo{date}{Jan.} \bibinfo{year}{2001}), \bibinfo{pages}{86}.
\newblock
\showISSN{1538-3873}
\urldef\tempurl%
\url{https://doi.org/10.1086/317967}
\showDOI{\tempurl}
\newblock
\shownote{Publisher: The University of Chicago Press}.


\bibitem[Witchel et~al\mbox{.}(2018)]%
        {witchel2018thighderived}
\bibfield{author}{\bibinfo{person}{Harry~J Witchel},
  \bibinfo{person}{C{\"a}cilia Oberndorfer}, \bibinfo{person}{Robert Needham},
  \bibinfo{person}{Aoife Healy}, \bibinfo{person}{Carina E~I Westling},
  \bibinfo{person}{Joseph~H Guppy}, \bibinfo{person}{Jake Bush},
  \bibinfo{person}{Jens Barth}, \bibinfo{person}{Chantal Herberz},
  \bibinfo{person}{Daniel Roggen}, \bibinfo{person}{Bj{\"o}rn~M Eskofier},
  \bibinfo{person}{Waqar Rashid}, \bibinfo{person}{Nachiappan Chockalingam},
  {and} \bibinfo{person}{Jochen Klucken}.} \bibinfo{year}{2018}\natexlab{}.
\newblock \showarticletitle{{Thigh-Derived} Inertial Sensor Metrics to Assess
  the {Sit-to-Stand} and {Stand-to-Sit} Transitions in the Timed Up and Go
  ({TUG}) Task for Quantifying Mobility Impairment in Multiple Sclerosis}.
\newblock \bibinfo{journal}{\emph{Front Neurol}}  \bibinfo{volume}{9}
  (\bibinfo{date}{Sept.} \bibinfo{year}{2018}), \bibinfo{pages}{684}.
\newblock


\bibitem[Zhang et~al\mbox{.}(2020a)]%
        {zhang2020videoacc}
\bibfield{author}{\bibinfo{person}{Yun~C. Zhang}, \bibinfo{person}{Shibo
  Zhang}, \bibinfo{person}{Miao Liu}, \bibinfo{person}{Elyse Daly},
  \bibinfo{person}{Samuel Battalio}, \bibinfo{person}{Santosh Kumar},
  \bibinfo{person}{Bonnie Spring}, \bibinfo{person}{James~M. Rehg}, {and}
  \bibinfo{person}{Nabil Alshurafa}.} \bibinfo{year}{2020}\natexlab{a}.
\newblock \showarticletitle{SyncWISE: Window Induced Shift Estimation for
  Synchronization of Video and Accelerometry from Wearable Sensors}.
\newblock \bibinfo{journal}{\emph{Proc. ACM Interact. Mob. Wearable Ubiquitous
  Technol.}} \bibinfo{volume}{4}, \bibinfo{number}{3}, Article
  \bibinfo{articleno}{107} (\bibinfo{date}{sep} \bibinfo{year}{2020}),
  \bibinfo{numpages}{26}~pages.
\newblock
\urldef\tempurl%
\url{https://doi.org/10.1145/3411824}
\showDOI{\tempurl}


\bibitem[Zhang et~al\mbox{.}(2020b)]%
        {zhang2020syncwise}
\bibfield{author}{\bibinfo{person}{Yun~C Zhang}, \bibinfo{person}{Shibo Zhang},
  \bibinfo{person}{Miao Liu}, \bibinfo{person}{Elyse Daly},
  \bibinfo{person}{Samuel Battalio}, \bibinfo{person}{Santosh Kumar},
  \bibinfo{person}{Bonnie Spring}, \bibinfo{person}{James~M Rehg}, {and}
  \bibinfo{person}{Nabil Alshurafa}.} \bibinfo{year}{2020}\natexlab{b}.
\newblock \showarticletitle{Syncwise: Window induced shift estimation for
  synchronization of video and accelerometry from wearable sensors}.
\newblock \bibinfo{journal}{\emph{Proceedings of the ACM on Interactive,
  Mobile, Wearable and Ubiquitous Technologies}} \bibinfo{volume}{4},
  \bibinfo{number}{3} (\bibinfo{year}{2020}), \bibinfo{pages}{1--26}.
\newblock


\bibitem[Zheng et~al\mbox{.}(2014)]%
        {Zheng2014unobtrusive}
\bibfield{author}{\bibinfo{person}{Ya-Li Zheng}, \bibinfo{person}{Xiao-Rong
  Ding}, \bibinfo{person}{Carmen Chung~Yan Poon}, \bibinfo{person}{Benny
  Ping~Lai Lo}, \bibinfo{person}{Heye Zhang}, \bibinfo{person}{Xiao-Lin Zhou},
  \bibinfo{person}{Guang-Zhong Yang}, \bibinfo{person}{Ni Zhao}, {and}
  \bibinfo{person}{Yuan-Ting Zhang}.} \bibinfo{year}{2014}\natexlab{}.
\newblock \showarticletitle{Unobtrusive sensing and wearable devices for health
  informatics}.
\newblock \bibinfo{journal}{\emph{IEEE Trans Biomed Eng}} \bibinfo{volume}{61},
  \bibinfo{number}{5} (\bibinfo{date}{May} \bibinfo{year}{2014}),
  \bibinfo{pages}{1538--1554}.
\newblock


\end{thebibliography}
\end{document}